\documentclass[11pt,a4paper]{article}
\usepackage[utf8]{inputenc}
\usepackage[margin=25mm]{geometry}
\usepackage{amsmath,amssymb,amsthm,mathtools}
\usepackage[shortlabels]{enumitem}
\usepackage[expansion=false]{microtype}
\usepackage{graphicx}
\usepackage{xcolor}
\usepackage[font=small,labelfont=bf]{caption}
\usepackage[colorlinks=true,linkcolor=blue!45!black,citecolor=blue!45!black,urlcolor=blue!45!black]{hyperref}
\graphicspath{{figures/}}

\numberwithin{equation}{section}
\newtheorem{theorem}{Theorem}[section]
\newtheorem{proposition}[theorem]{Proposition}
\newtheorem{lemma}[theorem]{Lemma}
\newtheorem{corollary}[theorem]{Corollary}
\theoremstyle{remark}
\newtheorem{remark}[theorem]{Remark}
\newtheorem{example}[theorem]{Example}

\newcommand{\Hil}{\mathcal H}
\newcommand{\Hphy}{\mathcal H_{\rm phy}}
\newcommand{\Ps}{\mathcal P}
\newcommand{\Pphy}{\mathcal P_{\rm phy}}
\newcommand{\R}{\mathbb R}

\newcommand{\Z}{\mathbb Z}
\newcommand{\CP}{\mathbb{C}P}
\newcommand{\PP}{\mathbb P}
\newcommand{\Mhat}{\widehat M}
\newcommand{\Chat}{\widehat C}
\newcommand{\eps}{\epsilon}
\newcommand{\ip}[2]{\langle #1|#2\rangle}
\newcommand{\ev}[1]{\langle #1\rangle}
\newcommand{\Op}{{\rm Op}_\hbar}
\newcommand{\Dmat}{\mathcal C}
\DeclareMathOperator{\Hess}{Hess}
\let\Im\relax
\DeclareMathOperator{\Im}{Im}
\DeclareMathOperator{\rank}{rank}
\DeclareMathOperator{\Tr}{Tr}
\DeclareMathOperator{\spec}{spec}
\DeclareMathOperator{\codim}{codim}

\title{\textbf{Recovering classical gauge geometry from quantum master constraints}}
\author{Alejandro Corichi\\[6pt]
\small Centro de Ciencias Matem\'aticas, Universidad Nacional Aut\'onoma de M\'exico,\\
\small UNAM-Campus Morelia, A.~Postal 61-3, Morelia, Michoac\'an 58190, Mexico\\
\small \texttt{corichi@matmor.unam.mx}}
\date{}

\begin{document}
\maketitle

\begin{abstract}
In constrained systems, from gauge theories to gravity, constraints do two jobs: they select the physical states and, classically, they generate gauge transformations. When all constraints are imposed through a single positive master operator, as in the master-constraint programme, it shows neither the gauge transformations nor whether the constraints were first or second class. We show where this information goes and how quantum theory recovers it. Near the constraint surface, the classical master dynamics turns small displacements off the surface into motion along the gauge orbits for first-class constraints, and into rotations for second-class ones; counting the two reproduces Dirac's classification. The same distinction controls the lowest eigenvalue of the master operator, the zero-point offset subtracted to find physical states. For first-class constraints we prove that this offset is at most of order $\hbar^2$, for any master constraint, quadratic or not. When second-class constraints are present everywhere it is of order $\hbar$, like the lowest Landau level of a charged particle in a magnetic field. This explains the $\hbar^2$ and $\hbar$ offsets found by Dittrich and Thiemann. A magnetic field that vanishes along a curve shows the limits of this test: the offset scales as $\hbar^{4/3}$, though the constraints are second class almost everywhere. Gauge directions also appear in semiclassical wave packets, whose fluctuations the master evolution shears along the gauge orbits. Two oscillators at fixed total energy are solved exactly. For the SL(2,$\R$) model with two Hamiltonian constraints we obtain the spectral bottom $\hbar^2/2$ and a propagation estimate.
\end{abstract}


\tableofcontents

\section{Introduction}\label{sec:intro}

In the Hamiltonian description of general relativity and of gauge theories, the constraints do much more than restrict the canonical variables. For a regular first-class system, their Hamiltonian vector fields span the degenerate directions of the symplectic form restricted to the constraint surface. Points on the same integral leaf cannot be distinguished by physical observables, and the physical phase space is the quotient along these leaves. This geometry makes precise the relation between constraints, gauge transformations and observables \cite{Dirac,GNH,HT,Gotay}. It is also at the heart of the way we think about the dynamics of general relativity, in which the evolution generated by the Hamiltonian constraint is itself a gauge motion.

The quantum description is different. In the Dirac implementation, physical states satisfy $\Chat_i\Psi=0$. When the solutions are normalizable, their rays form an embedded projective space, the induced symplectic form is nondegenerate, and the constraint-generated motions vanish at physical rays, so no quotient remains to be taken. This difference was analysed in \cite{Corichi2008} within the geometric formulation of quantum mechanics, in which pure states form a K\"ahler manifold and unitary transformations act by K\"ahler isometries \cite{Kibble,CMP,AshtekarSchilling,BrodyHughston}. The revised version of \cite{Corichi2008} explains why. The function on the space of rays that implements the Dirac conditions, the expectation value of the squared constraint, decomposes into the square of the expectation value, which mirrors the classical constraint, plus the squared speed of the quantum gauge orbit. The Dirac condition requires both to vanish. Rather than identifying the points of a gauge orbit, the quantum condition retains only the points whose gauge orbit has collapsed to a point. The quantum physical space must therefore be compared with the classical reduced phase space, not with the unreduced constraint surface.

This leaves the question addressed in this article. If quantum theory is fundamental, where do the unreduced classical surface, its degenerate directions and its first-class character come from? The question is sharpest when the constraints are imposed through a positive master operator $\Mhat$ \cite{Thiemann,DT1}, as in the master-constraint programme for loop quantum gravity. Such an operator combines all constraints into one condition. It commutes with itself, and the Hamiltonian vector field of its classical counterpart vanishes on the constraint surface. Neither fact displays the original gauge generators, nor whether the constraints were first or second class.

The master-constraint programme already provides a criterion for observables through a double Poisson bracket \cite{Thiemann,DT1}, and \cite{Corichi2008} gave a geometric interpretation of the corresponding double commutator. Its revised version shows that this condition is exact, that $\ev{\Mhat}$ is a Morse--Bott function on the space of rays whose transverse Hessian is the spectrum of $\Mhat$, and that along coherent states $\ev{\Mhat}$ equals the classical master function plus a term of order $\hbar$ given by the squared speed of the classical gauge orbits. Those statements concern the value of the master function and its Hessian. Here we examine the response of the master Hamiltonian vector field to displacements away from its zero set. A vector field can vanish on a submanifold while its linearization there has a nontrivial image. For a nonnegative master function this linearization is the \emph{Hamilton map}, or fundamental matrix, familiar from the microlocal analysis of operators with double characteristics \cite{Melin,BoutetdeMonvel,Sjostrand,Hormander85}. Its image is the symplectic orthogonal of the zero set, and its Jordan structure records how many constraints are first class and how many are second class. The image itself is fixed by the zero set and the symplectic form. What makes the structure useful for constraint quantization is that it is the derivative of the master dynamics, and that it controls the bottom of the master spectrum, so that quantum data can detect it without a prior choice of gauge generators.

The contributions of the paper are the following.
\begin{enumerate}[(i)]
\item \emph{Dirac's classification from the master dynamics} (Theorem~\ref{thm:reconstruction}). For a Morse--Bott master function, $\ker L=T\Sigma$ and $\Im L=(T\Sigma)^\omega$. The number of first-class constraints is $\rank L-\rank L^2$ and the number of second-class constraints is $\rank L^2$. The zero set is coisotropic exactly when $L^2=0$ and symplectic exactly when $L$ is semisimple. For quadratic masters, $L$ restricted to its image is the Dirac matrix of Poisson brackets weighted by the master weights (equation~(\ref{eq:LDirac})). In a K\"ahler frame, $L$ maps the imaginary gauge directions onto the real ones (Proposition~\ref{prop:kahler}).
\item \emph{The exact physical space} is the semisimple, rotational case of the same dichotomy (Proposition~\ref{prop:quantum}).
\item \emph{Spectral offsets.} For a nonnegative quadratic master, the bottom of the spectrum is $(\hbar/2)\Tr_+L$ (Proposition~\ref{prop:offset}), and for quadratic masters of constraints it saturates the central-term bound of \cite{Corichi2008} (Corollary~\ref{cor:dirac}). For general masters we prove two upper bounds with explicit trial states (Proposition~\ref{prop:general}): a first-class zero set gives a spectral bottom of order $\hbar^2$ at most, and at every point $z$ of the zero set the bottom is at most $(\hbar/2)\Tr_+L_z+O(\hbar^{4/3})$, with an exponent that cannot be improved. Combined with Melin's inequality, this explains the offsets of order $\hbar^2$ for first-class and of order $\hbar$ for second-class systems found by Dittrich and Thiemann \cite{DT2,DT3}. A magnetic example (Example~\ref{ex:magnetic}) shows the limit of this test: the offset detects the point of the zero set that is least second class, not the class of the constraints.
\item \emph{A fluctuation diagnostic.} The second time derivative of the limiting Gaussian covariance of a semiclassical packet has the gauge distribution as its image, even at a fixed packet centre, and it can be defined through bounded probes. Its trace and the $O(\hbar)$ offset of the master expectation are two moments of the same matrix (Remark~\ref{rem:moments}).
\item \emph{Two examples.} For two oscillators at fixed total energy we obtain the exact master evolution of coherent packets and their exact covariance, which realizes the gauge shear with error $O(\hbar)$ (Proposition~\ref{prop:osc-cov}), together with the reduced symplectic limit of projected physical states. For the SL(2,$\R$) model of \cite{MRT} we compute the coherent and Weyl master symbols, prove essential self-adjointness of the master operator and a finite-time propagation estimate, and derive the spectral bottom $\hbar^2/2$ from the su(1,1) structure.
\end{enumerate}

Several ingredients are known, and we identify them as such. The linear algebra behind Theorem~\ref{thm:reconstruction} and Proposition~\ref{prop:offset} is that of the Hamilton map of a nonnegative symbol at a double characteristic, and the distinction between symplectic and involutive double characteristics, the positive trace $\Tr_+$ and Melin's lower bound are classical in microlocal analysis \cite{Melin,BoutetdeMonvel,Sjostrand,Hormander85}. The double-bracket criterion \cite{Thiemann,Corichi2008}, the projected oscillator states \cite{ABC}, Gaussian wave-packet propagation \cite{Heller,Hagedorn,Littlejohn,CombescureRobert} and the normal form of nonnegative quadratic forms \cite{Williamson,Hormander85,Hormander95} are also established. What is new is their use as a diagnostic for constraint quantization: the reading of Dirac's classification off the master dynamics, the upper bounds on the spectral offset of general master operators, the identification of what the order of the offset can and cannot detect, and the semiclassical and model-specific realizations. The SL(2,$\R$) model has also been quantized by group averaging, by the master constraint and by uniform discretizations \cite{LoukoRovelli,DT3,GambiniOlmedo}; we keep the choice of physical quantum sector separate from the reconstruction of the gauge geometry.

The geometry of the group-averaging map, which produces physical states from kinematical seeds, is analysed in \cite{CorichiGA}, including its extension to distributional physical states. The same question is addressed without any master function in \cite{CorichiDirac}, where the classical gauge geometry is recovered from the physical projector itself: the probability that a kinematical state is physical is constant along the gauge orbits and, for a compact gauge group, is maximal within each complexified gauge orbit exactly where the constraint expectations vanish. These articles can be read independently. They meet at a precise point, which we discuss in Section~\ref{sec:realimag}: the complexified gauge directions that group averaging discards are the directions that the linearized master flow relates, and the probability of being physical is what locates the constraint surface along them.

Section~\ref{sec:physics} gives a nontechnical account of the physical mechanisms behind the results, which can be read on its own. Section~\ref{sec:geometries} recalls the classical and quantum geometries to be compared. Section~\ref{sec:reconstruction} contains the structure theorem and the spectral results. Section~\ref{sec:semiclassical} formulates the semiclassical requirements and the fluctuation diagnostic. Sections~\ref{sec:oscillator} and \ref{sec:sl2r} treat the two examples, and Section~\ref{sec:discussion} discusses the implications, including what is needed for gravity. Section~\ref{sec:conclusions} concludes. Proofs and derivations are collected in the appendices.

\section{The physical picture}\label{sec:physics}

This section describes, in physical terms and without proofs, the mechanisms behind the results. Readers interested only in the physics can read it together with the examples of Sections~\ref{sec:oscillator} and~\ref{sec:sl2r} and the discussion of Section~\ref{sec:discussion}. The precise statements, hypotheses and proofs are in Sections~\ref{sec:reconstruction} and~\ref{sec:semiclassical}.

\subsection{The master flow is constraint flow with self-generated lapse and shift}

Write the master function as $m=\frac12K^{ij}C_iC_j$, with positive constant weights $K$, as in (\ref{eq:quadmaster}). Its Hamiltonian vector field is
\begin{equation}\label{eq:Xmphys}
X_m=K^{ij}\,C_j\,X_{C_i} .
\end{equation}
It vanishes on the constraint surface $\Sigma$, so every point of $\Sigma$ is a fixed point of the master dynamics, which therefore seems to contain no gauge motion at all. At a nearby point, where the constraints take small values $c_i=C_i(z)$, (\ref{eq:Xmphys}) says that the point moves along the constraint flows with ``lapse and shift'' $Kc$, set by its own constraint values. These values evolve as well, exactly, by
\begin{equation}\label{eq:cdotphys}
\dot c=\Dmat\,K\,c,\qquad \Dmat_{ij}=\{C_i,C_j\},
\end{equation}
with $\Dmat$ evaluated along the trajectory. For first-class constraints $\Dmat$ vanishes on $\Sigma$, so the constraint values are constant to first order and the point drifts along a gauge orbit with constant velocity: a displacement off the surface is converted into motion along the gauge orbits. For second-class constraints $K\Dmat$ has imaginary eigenvalues $\pm i\mu_j$, the constraint values rotate with angular frequencies $\mu_j$, and the point oscillates about the surface without drifting (Figure~\ref{fig:shear}). Counting the two behaviours reproduces Dirac's classification, without a prior choice of gauge generators (Theorem~\ref{thm:reconstruction}). For a Hamiltonian constraint $C=H-E$, whose gauge orbits are the classical trajectories, (\ref{eq:Xmphys}) is ordinary time evolution with a lapse equal to the energy mismatch $H-E$: trajectories slightly off the energy shell evolve at a rate proportional to how far off the shell they are.

\subsection{Why second-class constraints cost a zero-point energy and first-class ones do not}

Quantum mechanically the same distinction controls the bottom of the master spectrum. Two second-class constraints with constant bracket $\{C_1,C_2\}=b$ satisfy $[\Chat_1,\Chat_2]=i\hbar b$, so by the uncertainty relation they cannot vanish sharply together, and
\begin{equation}\label{eq:uncertphys}
\ev{\Chat_1^2}+\ev{\Chat_2^2}\ \ge\ 2\,\Delta C_1\,\Delta C_2\ \ge\ \hbar|b| .
\end{equation}
In fact $\frac12(\Chat_1^2+\Chat_2^2)$ is a harmonic oscillator of frequency $|b|$, which is the classical rotation frequency of (\ref{eq:cdotphys}). Its ground states, which saturate (\ref{eq:uncertphys}), are the Gupta--Bleuler states (Corollary~\ref{cor:dirac}). In general each second-class pair contributes the zero-point energy $\hbar\mu_j/2$ of an oscillator. First-class constraints commute on $\Sigma$ and can be made sharp together: a state can be spread along the gauge orbits with zero conjugate momentum, like a plane wave of zero momentum, and normalizable approximations to it cost only ordering and curvature effects of order $\hbar^2$. Proposition~\ref{prop:general} turns this picture into an upper bound valid for master operators of any form, and Melin's inequality gives the matching lower bound of order $\hbar$ when second-class constraints are present everywhere. This is the origin of the offsets of order $\hbar^2$ and $\hbar$ observed by Dittrich and Thiemann \cite{DT2,DT3}.

A charged particle in a plane makes the picture concrete (Example~\ref{ex:magnetic}). The kinetic momenta $\xi_i-A_i$ are a second-class pair with bracket equal to the magnetic field, the master operator is the kinetic energy, and its offset is the lowest Landau level, $\hbar|B|/2$ for a uniform field. Subtracting it and keeping the lowest level is the familiar reduction to the guiding-centre dynamics. The same example shows what the offset cannot see. It is a global minimum, set by the part of $\Sigma$ that is \emph{least} second class: a field that vanishes along a line gives an offset of order $\hbar^{4/3}$, although the constraints are second class almost everywhere. An offset of order $\hbar$ thus reveals second-class constraints everywhere, while a smaller offset does not prove that the constraints are first class. Table~\ref{tab:offsetsphys} collects the examples of this paper.

\begin{table}[t]
\centering\small
\renewcommand{\arraystretch}{1.2}
\begin{tabular}{@{}llll@{}}
\hline
System & Constraints & Class & Offset $\inf\spec\Mhat$\\
\hline
Free particle & $p$ & first & $0$\\
Two oscillators at fixed energy (Section~\ref{sec:oscillator}) & $H-E$ & first & $\le\hbar^2\nu^2/8$\\
SL(2,$\R$) model (Section~\ref{sec:sl2r}) & $H_1,H_2,D$ & first & $\hbar^2/2$\\
One canonical pair & $q,\ p$ & second & $\hbar/2$\\
Charge in a uniform field $B$ & $\xi-A$ & second & $\hbar|B|/2$\\
Charge in the field $B=1+x_1^2$ & $\xi-A$ & second & $\hbar/2+o(\hbar)$\\
Charge in the field $B=x_1$ & $\xi-A$ & second off a line & $0.285\,\hbar^{4/3}$\\
\hline
\end{tabular}
\caption{Zero-point offsets of the master operators of the examples, with $m=\frac12\sum_iC_i^2$ except for the SL(2,$\R$) model, whose master is given in Section~\ref{sec:sl2r}.}
\label{tab:offsetsphys}
\end{table}

\subsection{Wave packets spread along the gauge orbits}

The spectrum needs only the master operator. Semiclassical states see more. A coherent state centred on $\Sigma$ has vanishing mean constraints but nonvanishing fluctuations, of order $\hbar^{1/2}$; the fluctuation of a constraint is the speed with which it moves the state, the quantum remnant of its gauge orbit \cite{Corichi2008}. Under the master unitary the centre stays fixed, since points of $\Sigma$ are fixed points of the classical master flow, but the components of the packet with constraint values of order $\hbar^{1/2}$ drift along the gauge orbits with velocities proportional to those values. The packet therefore stays centred on $\Sigma$ and spreads along its gauge orbits, linearly in the master parameter, while its widths across $\Sigma$ stay fixed (Section~\ref{sec:semiclassical}). For second-class constraints the covariance rotates instead. For a Hamiltonian constraint the spreading is dephasing along the packet's own classical trajectory: components with energy $E+\delta E$, with $\delta E\sim(\hbar\nu E)^{1/2}$ for the oscillators of Section~\ref{sec:oscillator}, move along the trajectory at rates proportional to $\delta E$. The static offset of a coherent state and the rate at which it spreads are the first and second moments of one matrix (Remark~\ref{rem:moments}).

In all three forms the gauge structure is visible only near the constraint surface, in kinematical rather than physical states, and through the response of the master dynamics. This is the sense in which classical gauge orbits are recovered from quantum master constraints.

\section{The geometries of classical and quantum reduction}\label{sec:geometries}

We begin by recalling the two geometries to be compared. The classical one is standard; the quantum one is the geometric formulation of quantum mechanics, in the normalization of \cite{AshtekarSchilling,Corichi2008}.

\subsection{Classical constraints and degenerate directions}

Let $(\Gamma,\omega)$ be a finite-dimensional symplectic manifold. We use
\begin{equation}\label{eq:conventions}
\iota_{X_f}\omega={\rm d}f,\qquad \{f,g\}={\rm d}f(X_g).
\end{equation}
For $\omega=\sum_a{\rm d}q^a\wedge{\rm d}p_a$ these conventions give $\{q^a,p_b\}=\delta^a_b$, $X_f=(\partial_pf,-\partial_qf)$ and
\begin{equation}\label{eq:bracket}
[X_f,X_g]=-X_{\{f,g\}} .
\end{equation}
We write $\omega^\flat(\delta z)=\iota_{\delta z}\omega$ and $\Pi^\sharp=(\omega^\flat)^{-1}$. In canonical coordinates $\Pi^\sharp$ is the matrix $\Pi=\bigl(\begin{smallmatrix}0&1\\-1&0\end{smallmatrix}\bigr)$ acting on gradients, so that $X_f=\Pi\nabla f$. We refer to Hamiltonian vector fields as HVFs.

For an embedded submanifold $\Sigma\subset\Gamma$ with restricted form $\omega_\Sigma=\iota^*\omega$, the characteristic distribution is
\begin{equation}\label{eq:characteristic}
D_z=\ker(\omega_\Sigma)_z=T_z\Sigma\cap(T_z\Sigma)^\omega .
\end{equation}
$\Sigma$ is coisotropic when $(T_z\Sigma)^\omega\subset T_z\Sigma$, and then $D_z=(T_z\Sigma)^\omega$; it is symplectic when $D_z=0$. For a regular first-class system with independent defining functions $C_i$, $D={\rm span}\{X_{C_i}\}$. The constraint flows are symplectomorphisms and their leaves in $\Sigma$ are isotropic. If the leaf space is smooth and the quotient map $\pi:\Sigma\to\Gamma_{\rm red}$ is a submersion, there is a unique symplectic form with
\begin{equation}\label{eq:reduced}
\pi^*\omega_{\rm red}=\omega_\Sigma .
\end{equation}
Global smoothness and Hausdorffness of the quotient are additional assumptions, which matter for the noncompact example of Section~\ref{sec:sl2r}. For a general regular constraint set, Dirac's classification is pointwise linear algebra: at $z\in\Sigma$, $\dim D_z$ independent combinations of the constraints are first class, and the remaining $\codim\Sigma-\dim D_z$, an even number, are second class \cite{Dirac,HT}.

Coisotropy is a property of the embedded submanifold. Any local defining functions of a coisotropic $\Sigma$ satisfy $\{C_i,C_j\}|_\Sigma=0$, and conversely this property for one set of defining functions implies coisotropy. Moreover, the coisotropic normal form provides local defining functions that Poisson-commute identically in a neighbourhood \cite{Gotay}. Structure functions therefore depend on the chosen generators, while the characteristic distribution does not. The aim of reconstruction is first to recover the submanifold, its characteristic distribution and the class of the constraints; a basis of gauge generators, generally only local, comes afterwards.

\subsection{Quantum states as a symplectic manifold}

Let $\Hil$ be a complex Hilbert space and $\Ps=\PP(\Hil)$ its space of rays $[\Psi]$. The Hermitian inner product induces a metric $g$, a complex structure $J$ and a symplectic form $\Omega$ on $\Ps$ \cite{Kibble,CMP,AshtekarSchilling,BrodyHughston}. With $\hbar$ explicit, our normalization is
\begin{equation}\label{eq:Omega}
\Omega=i\hbar\,\partial\bar\partial\log\|\Psi\|^2 ,
\end{equation}
so that $\ip{\Psi}{\Phi}=(2\hbar)^{-1}(g+i\Omega)$ on horizontal vectors at a normalized representative. A self-adjoint operator $\widehat A$ defines the function $a([\Psi])=\ip{\Psi}{\widehat A\Psi}/\ip{\Psi}{\Psi}$, whose Hamiltonian flow is the projectivization of $e^{-it\widehat A/\hbar}$. At a normalized representative, the horizontal HVF and the Poisson bracket are
\begin{equation}\label{eq:quantumHVF}
X_a=-\frac1\hbar\,J(\widehat A-a)\Psi,\qquad \{a,b\}_\Ps=\frac{1}{i\hbar}\ev{[\widehat A,\widehat B]} .
\end{equation}
The metric norm of $X_a$ is proportional to the dispersion of $\widehat A$, so $X_a$ vanishes exactly at eigenrays.

The condition $\ev{\Chat}=0$ is not equivalent to $\Chat\Psi=0$, since positive and negative spectral contributions can cancel. By contrast, for a normalized $\Psi$ in the domain of $\Chat$,
\begin{equation}\label{eq:squared}
\ev{\Chat^2}=\|\Chat\Psi\|^2=0\iff\Chat\Psi=0 ,
\end{equation}
and a positive sum of such squares selects the common kernel. The geometric function implementing the exact quantum condition is therefore the expectation of a squared operator \cite{Corichi2008}. Its excess over $\ev{\Chat}^2$ is the dispersion $(\Delta\Chat)^2$, the squared speed of the quantum gauge orbit, and it is this term that has no classical counterpart on the constraint surface. The function $M=\ev{\Mhat}$ used below is the function denoted $\mathcal C$ in \cite{Corichi2008,CorichiGA}.

\subsection{The embedded physical space}

Suppose that $\Hphy=\ker\Mhat$ is a closed nonzero subspace, for a self-adjoint $\Mhat\ge0$. Its projectivization $\Pphy$ is a complex submanifold, so (\ref{eq:Omega}) restricts to a symplectic form on it. The constraint unitaries act trivially on $\Pphy$, and physical observables act within it, so no additional quotient is needed. The following statement places this fact within the classification of Section~\ref{sec:reconstruction}.

\begin{proposition}[Quantum physical space]\label{prop:quantum}
Let $u\in\Hphy$ be a unit vector, and suppose that $0$ is an isolated point of the spectrum of $\Mhat$, with gap $\eta>0$ (this is automatic if $\dim\Hil<\infty$). For horizontal variations $\delta\Psi\perp u$ in the form domain of $\Mhat$, the function $M=\ev{\Mhat}$ satisfies
\begin{equation}\label{eq:quantumHess}
\Hess_uM(\delta\Psi,\delta\Psi)=2\ip{\delta\Psi}{\Mhat\delta\Psi},
\end{equation}
which vanishes exactly on $T_u\Pphy$ and is bounded below by $2\eta\|\delta\Psi\|^2$ on its orthogonal complement. The linearization of $X_M$ at $u$ is
\begin{equation}\label{eq:LQ}
L^Q_u\,\delta\Psi=-\frac1\hbar\,J\Mhat\,\delta\Psi ,
\end{equation}
with kernel $T_u\Pphy$ and image the symplectic normal space, on which $(L^Q_u)^2=-\Mhat^2/\hbar^2$ is negative definite. Thus $L^Q_u$ acts on the normal space as a rotation, with frequencies in $\hbar^{-1}(\spec\Mhat\setminus\{0\})$, and $\Im L^Q_u\cap\ker L^Q_u=\{0\}$.
\end{proposition}

\begin{proof}
Since $\Mhat u=0$, $M(u+\eps\delta\Psi)=\eps^2\ip{\delta\Psi}{\Mhat\delta\Psi}/(1+\eps^2\|\delta\Psi\|^2)$, which gives (\ref{eq:quantumHess}); the gap gives the lower bound. Linearizing $X_M=-\hbar^{-1}J(\Mhat-M)\Psi$ at $u$, the variation of $M$ vanishes and $M(u)=0$, which gives (\ref{eq:LQ}). The vector $\Mhat\delta\Psi$ is orthogonal to $u$, and $J$ commutes with the complex-linear $\Mhat$. Because $\Pphy$ is a complex submanifold, its metric and symplectic normal spaces coincide, and $\Mhat$ is invertible on the normal space by the gap.
\end{proof}

The identity (\ref{eq:quantumHess}) and the description of the normal Hessian by the spectrum of $\Mhat$ are in \cite{Corichi2008}, where they are used to show that $M$ is a Morse--Bott function with $\Pphy$ as its minimum. What we add is the linearization (\ref{eq:LQ}) and its square. In the language of Theorem~\ref{thm:reconstruction} below, the normal response at an exact physical ray is a pure rotation, as for a second-class zero set, never a shear into gauge directions. This is a restatement, not a new reason, of the fact that $\Pphy$ is symplectic: being a complex submanifold already implies it. Without a gap the Hessian is not uniformly nondegenerate on the normal space, and the identification of the image requires a closure.

The classical object corresponding to $\Pphy$ is therefore $\Gamma_{\rm red}$, and the unreduced surface $\Sigma$ must emerge from a different comparison, involving kinematical semiclassical states. When physical solutions are distributional, as in the noncompact model below, a rigging map and a physical inner product replace the embedded-kernel description \cite{ALMMT,GiuliniMarolf,LoukoRovelli}; the geometry of that construction is the subject of \cite{CorichiGA}.

\section{The Hamilton map of a master function}\label{sec:reconstruction}

A master function replaces a list of constraints that vanish transversally by a single function that vanishes quadratically. Its first derivative vanishes on the zero set, which is why its HVF vanishes there, but its second derivative does not. We now examine that second derivative before imposing any gauge interpretation.

Let $m\ge0$ be smooth, with zero set $\Sigma=m^{-1}(0)$ an embedded submanifold. Every point of $\Sigma$ is critical, so $X_m|_\Sigma=0$. At $z\in\Sigma$ the Hessian $\Hess_zm$ and the linearization
\begin{equation}\label{eq:L}
L_z={\rm d}X_m|_z:T_z\Gamma\to T_z\Gamma
\end{equation}
are intrinsic: changes of coordinates or connection contribute terms proportional to ${\rm d}m$ or $X_m$, which vanish at $z$. Since $z$ is a minimum, $\Hess_zm\ge0$. The hypothesis of the next theorem is that $\Hess_zm$ is nondegenerate on the normal space, i.e.\ that $\Sigma$ is a Morse--Bott critical manifold \cite{Bott}. Since $\Sigma$ is not a regular level set of $m$, its regularity is an independent assumption.

\begin{theorem}[Hamilton map of a master function]\label{thm:reconstruction}
Assume $\ker(\Hess_zm)^\flat=T_z\Sigma$ at every $z\in\Sigma$. Then:
\begin{enumerate}[(a)]
\item $L_z=\Pi^\sharp_z\,(\Hess_zm)^\flat$, with
\begin{equation}\label{eq:reconb}
\ker L_z=T_z\Sigma,\qquad \Im L_z=(T_z\Sigma)^\omega,\qquad \rank L_z=\codim\Sigma .
\end{equation}
\item $\Im L_z\cap\ker L_z=D_z$, and
\begin{equation}\label{eq:counting}
\dim D_z=\rank L_z-\rank L_z^2,\qquad \codim\Sigma-\dim D_z=\rank L_z^2 .
\end{equation}
\item The spectrum of $L_z$ lies on the imaginary axis. Its nonzero eigenvalues $\pm i\mu_j$, $\mu_j>0$, $j=1,\dots,\frac12\rank L_z^2$, are semisimple, and the eigenvalue $0$ has exactly $\dim D_z$ Jordan blocks of size two and no larger blocks.
\item The following equivalences hold:
\begin{equation}\label{eq:squarezero}
\begin{gathered}
\Sigma\ \text{coisotropic}\iff L_z^2=0\ \ \forall z\iff L_z\ \text{nilpotent}\ \ \forall z,\\
\Sigma\ \text{symplectic}\iff L_z\ \text{semisimple}\ \ \forall z .
\end{gathered}
\end{equation}
When $D$ has constant rank, it is a smooth integrable distribution on $\Sigma$.
\end{enumerate}
\end{theorem}

\begin{proof}
Differentiating $\omega^\flat X_m={\rm d}m$ at $z$ gives $\omega^\flat_zL_z=(\Hess_zm)^\flat$, because the term containing the derivative of $\omega$ multiplies $X_m(z)=0$. This proves the formula for $L_z$ and the kernel statement. The range of a symmetric map is the annihilator of its kernel, so $\Im(\Hess_zm)^\flat=N^*_z\Sigma$ and $\Im L_z=\Pi^\sharp_zN^*_z\Sigma=(T_z\Sigma)^\omega$, whose dimension is the codimension of $\Sigma$. Part (b) follows: $\Im L_z\cap\ker L_z=(T_z\Sigma)^\omega\cap T_z\Sigma$, and applying rank--nullity to $L_z$ restricted to $\Im L_z$ gives $\rank L_z^2=\rank L_z-\dim(\Im L_z\cap\ker L_z)$. Part (c) follows from the symplectic normal form of nonnegative quadratic forms, (\ref{eq:williamson}) below, applied to $\Hess_zm$: in the normal form $L_z$ is a direct sum of rotation blocks with eigenvalues $\pm i\mu_j$, of nilpotent blocks $(\delta x,\delta\xi)\mapsto(\delta\xi,0)$, and of zero. The nilpotent blocks span $\Im L_z\cap\ker L_z$, so their number is $\dim D_z$. Part (d) restates (b) and (c). Smoothness and integrability of a constant-rank kernel of the closed form $\omega_\Sigma$ are shown in Appendix~\ref{app:reconstruction}.
\end{proof}

In Dirac's terms, (\ref{eq:counting}) says that at every point of $\Sigma$ the master dynamics counts the first-class constraints, $\rank L_z-\rank L_z^2$, and the second-class constraints, $\rank L_z^2$, without reference to a choice of generators. The image $(T_z\Sigma)^\omega$ and the counting are determined by $\Sigma$ and $\omega$; the Hessian enters through the hypothesis on its kernel and through the frequencies $\mu_j$, which depend on the weights. The content of the theorem is that this information is carried by the derivative of the master flow. Since $z$ is a fixed point, ${\rm d}\Phi^t_m|_z=e^{tL_z}$, and for a coisotropic zero set
\begin{equation}\label{eq:shear}
{\rm d}\Phi^t_m|_z=1+tL_z .
\end{equation}
A transverse displacement is sheared into a gauge direction, and a displacement tangent to $\Sigma$ is fixed to first order. The master flow fixes each point of $\Sigma$; it does not move a point along its gauge leaf. It is the response of nearby initial data that identifies the null directions of $\omega_\Sigma$, and this response can be read off from local dynamical data at a single point. Second-class directions respond instead by rotation, with frequencies $\mu_j$.

\begin{remark}[Relation to microlocal analysis]\label{rem:microlocal}
Up to normalization, $L_z$ is the Hamilton map, or fundamental matrix, of the nonnegative symbol $m$ at the doubly characteristic point $z$ \cite[section~21.5]{Hormander85}. The distinction between symplectic and involutive double characteristics, the role of the positive trace $\Tr_+L_z=\sum_j\mu_j$, and lower bounds of the form $\Op(m)\gtrsim\frac\hbar2\Tr_+L$ near $\Sigma$ go back to the study of hypoelliptic operators with double characteristics \cite{Melin,BoutetdeMonvel,Sjostrand} (see also \cite[section~22.3]{Hormander85}). Theorem~\ref{thm:reconstruction} is the same linear algebra, phrased for constraint systems. In that setting, the involutive case is the first-class case, and the symplectic case is the second-class case.
\end{remark}

\subsection{A local normal form}

\begin{figure}[t]
\centering
\includegraphics[width=\textwidth]{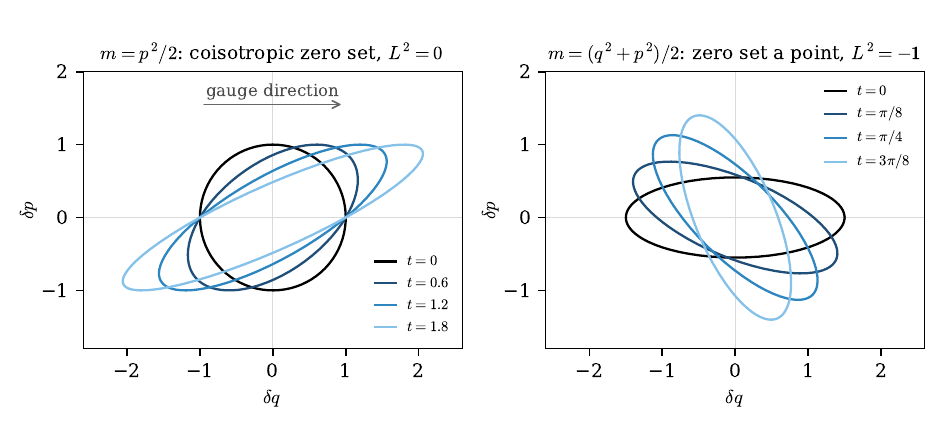}
\caption{Linearized master response in a single canonical plane. Left: for $m=p^2/2$, the zero set $p=0$ is coisotropic and the flow maps $(\delta q,\delta p)$ to $(\delta q+t\,\delta p,\delta p)$; a circular covariance contour is sheared along the gauge direction. Right: for $m=(q^2+p^2)/2$, the zero set is a point and an anisotropic contour rotates. The curves display the linear maps acting on covariance contours. The distinction is $L^2=0$ versus $L^2=-\mathbf 1$. Exact quantum covariances realizing the left panel are shown in Figure~\ref{fig:osc}.}
\label{fig:shear}
\end{figure}

For a coisotropic $\Sigma$ of codimension $r$, there are local Darboux coordinates $(Q^i,P_i,x^A,y_A)$, $i=1,\dots,r$, in which $\Sigma=\{P=0\}$ \cite{Gotay}. The $Q^i$ are gauge coordinates, the $P_i$ are transverse, and $(x,y)$ coordinatize the local reduced space. Since $m$ and ${\rm d}m$ vanish on $\Sigma$,
\begin{equation}\label{eq:normalform}
m=\tfrac12\kappa^{ij}(Q,x,y)\,P_iP_j+O(|P|^3),
\end{equation}
with $\kappa^{ij}$ positive definite by the Morse--Bott hypothesis. At $P=0$ the linearized field is
\begin{equation}\label{eq:normalL}
L(\delta Q,\delta P,\delta x,\delta y)=(\kappa\,\delta P,\,0,\,0,\,0).
\end{equation}
A transverse displacement $\delta P$ is mapped into a gauge displacement $\delta Q$, while physical tangent displacements and displacements along the gauge leaf are annihilated. Figure~\ref{fig:shear} contrasts this shear with the rotation produced by a second-class zero set.

\subsection{Positive quadratic master functions}

For independent constraints $C_i$, let
\begin{equation}\label{eq:quadmaster}
m=\tfrac12K^{ij}C_iC_j,\qquad K^{ij}=K^{ji}\ \text{constant and positive definite}.
\end{equation}
On $\Sigma=\{C_i=0\}$,
\begin{equation}\label{eq:quadL}
\Hess_zm=K^{ij}\,{\rm d}C_i|_z\otimes{\rm d}C_j|_z,\qquad L_z(\delta z)=K^{ij}\,{\rm d}C_j(\delta z)\,X_{C_i}(z).
\end{equation}
Independence and positivity give the Morse--Bott hypothesis, and the image of $L_z$ is spanned by the $X_{C_i}(z)$. Evaluating (\ref{eq:quadL}) on this frame gives
\begin{equation}\label{eq:LDirac}
L_z\bigl(X_{C_k}\bigr)=\sum_i(K\Dmat)_{ik}\,X_{C_i}(z),\qquad \Dmat_{jk}=\{C_j,C_k\}(z),
\end{equation}
so $L_z$ restricted to its image is represented by $K\Dmat$, the Dirac matrix of the constraints weighted by the master weights. Since $K\Dmat$ is similar to the antisymmetric matrix $K^{1/2}\Dmat K^{1/2}$, its eigenvalues are imaginary and semisimple, $\rank L_z^2=\rank\Dmat$ is Dirac's count of second-class constraints, and
\begin{equation}\label{eq:trKC}
\Tr_+L_z=\Tr_+(K\Dmat):=\tfrac12\sum_i|\lambda_i(K\Dmat)| .
\end{equation}
The image and the counting are independent of the weights; the frequencies are not. On a regular surface the first-class condition is exactly $L^2=0$.

The Morse--Bott hypothesis cannot be omitted. The function $m=p^4$ has the same coisotropic zero set as $p^2/2$, but its Hessian and $L$ vanish there, and higher derivatives would be needed. Likewise, at a singular stratum of a quadratic constraint system the gradients lose rank and the hypothesis fails. Positivity alone, on the other hand, does not imply first-classness. For $m=(q^2+p^2)/2$ on $\R^2$, $\Sigma$ is a point and $L(\delta q,\delta p)=(\delta p,-\delta q)$, with $L^2=-\mathbf 1$: a transverse rotation, characteristic of the second-class pair $(q,p)$. For $m=p^2/2$, $L(\delta q,\delta p)=(\delta p,0)$ squares to zero, and its image is the gauge direction of $p=0$.

When phase space carries a complex structure $J$ compatible with the symplectic form, as it does whenever it is parametrized by coherent states, the transverse directions acquire a further meaning. Let $g$ be the metric defined by $\omega(X,Y)=g(JX,Y)$. The gradient of a function is then $\nabla f=JX_f$, and the metric normal space of $\Sigma$ is $J$ applied to its symplectic orthogonal:
\begin{equation}\label{eq:normalJ}
(T_z\Sigma)^\perp=J\,(T_z\Sigma)^\omega .
\end{equation}
In the language of group actions, $X_{C_i}$ are the infinitesimal gauge motions and $JX_{C_i}$ their imaginary counterparts, the directions generated by the complexified gauge group.

\begin{proposition}\label{prop:kahler}
For the quadratic master (\ref{eq:quadmaster}) on a K\"ahler manifold $(\Gamma,\omega,g,J)$, let $G_{ij}=g(X_{C_i},X_{C_j})$ be the Gram matrix of the gauge fields at $z\in\Sigma$. Then
\begin{equation}\label{eq:LJ}
L_z\bigl(JX_{C_k}\bigr)=\sum_i(KG)_{ik}\,X_{C_i}(z).
\end{equation}
The linearized master flow maps the imaginary gauge directions onto the real ones, with the positive matrix $KG$.
\end{proposition}

\begin{proof}
By (\ref{eq:quadL}), $L_z(JX_{C_k})=K^{ij}\,{\rm d}C_j(JX_{C_k})\,X_{C_i}$, and ${\rm d}C_j(JX_{C_k})=g(\nabla C_j,\nabla C_k)=g(X_{C_j},X_{C_k})$, since $J$ is an isometry.
\end{proof}

The proposition is elementary, but it is the precise statement behind the shear picture: the master flow does not generate gauge motions; it converts imaginary gauge displacements, which move a point off the constraint surface, into real ones. We shall see in Section~\ref{sec:realimag} that the imaginary gauge directions are also the ones that group averaging discards.

\subsection{Recovering an algebra and its observables}

On a regular coisotropic surface choose independent local defining functions $C_i$. Then $\{C_i,C_j\}|_\Sigma=0$, and Hadamard's lemma gives smooth coefficients $F_{ij}{}^k$ with $\{C_i,C_j\}=F_{ij}{}^kC_k$ near $\Sigma$. By (\ref{eq:bracket}),
\begin{equation}\label{eq:algebra}
[X_{C_i},X_{C_j}]|_\Sigma=-F_{ij}{}^kX_{C_k}|_\Sigma .
\end{equation}
The distribution is fixed, but the coefficients depend on the defining functions. Replacing $C_i$ by $A_i{}^jC_j$, with $A$ invertible, changes the frame and adds derivative terms to the brackets. Reconstruction therefore yields a first-class algebra in any chosen local frame, not a preferred off-surface presentation. More intrinsically, the vanishing ideal $I_\Sigma$ satisfies $\{I_\Sigma,I_\Sigma\}\subset I_\Sigma$ and $\{I_\Sigma^2,I_\Sigma\}\subset I_\Sigma^2$. The quotient $I_\Sigma/I_\Sigma^2$, the space of sections of the conormal bundle, inherits a bracket that makes $N^*\Sigma$ a Lie algebroid \cite{Weinstein}; with the convention (\ref{eq:conventions}) its anchor is $[f]\mapsto-X_f|_\Sigma$. The same ideal underlies algebraic reduction \cite{SniatyckiWeinstein}. The master function lies in $I_\Sigma^2$, and its Hessian recovers the conormal module geometrically.

For a smooth observable $O$, (\ref{eq:quadmaster}) gives
\begin{equation}\label{eq:doublebracket}
\{O,\{O,m\}\}|_\Sigma=K^{ij}\{O,C_i\}\{O,C_j\}|_\Sigma=\Hess m(X_O,X_O)|_\Sigma .
\end{equation}
The double bracket therefore tests whether $X_O$ is tangent to $\Sigma$, equivalently whether $O|_\Sigma$ is constant on the null leaves, in which case $O$ descends to the quotient. This is the master-constraint criterion for observables \cite{Thiemann,DT1,Corichi2008}, evaluated through the same transverse Hessian that determines the gauge image.

\subsection{Spectral offsets}\label{sec:offset}

The classification of Theorem~\ref{thm:reconstruction} has a spectral counterpart. We use the Weyl quantization $\Op$ on $L^2(\R^n)$ in canonical coordinates.

\begin{proposition}[Quadratic masters]\label{prop:offset}
Let $q\ge0$ be a quadratic form on $(\R^{2n},\omega)$ and $L_q=\Pi\Hess q$. The spectrum of $L_q$ lies on the imaginary axis; let $\pm i\mu_j$, $\mu_j>0$, $j=1,\dots,r$, be its nonzero eigenvalues counted with multiplicity, and $\Tr_+L_q=\sum_j\mu_j$. Then
\begin{equation}\label{eq:offset}
\inf\spec\Op(q)=\frac\hbar2\Tr_+L_q ,
\end{equation}
and the following are equivalent: $L_q$ is nilpotent; $L_q^2=0$; the zero set $\ker\Hess q$ of $q$ is coisotropic; $\inf\spec\Op(q)=0$.
\end{proposition}

\begin{proof}
By the symplectic normal form of nonnegative quadratic forms \cite{Williamson,Hormander85,Hormander95}, there are linear Darboux coordinates in which
\begin{equation}\label{eq:williamson}
q=\sum_{j=1}^r\frac{\mu_j}2\bigl(x_j^2+\xi_j^2\bigr)+\sum_{k=r+1}^{r+s}\frac12\xi_k^2 .
\end{equation}
In these coordinates $L_q$ is a direct sum of rotations with eigenvalues $\pm i\mu_j$ and of blocks $(\delta x_k,\delta\xi_k)\mapsto(\delta\xi_k,0)$, which square to zero. Weyl quantization is exactly covariant under the metaplectic representation \cite{Folland}, so $\Op(q)$ is unitarily equivalent to a sum of $r$ independent oscillators with frequencies $\mu_j$ and $s$ free particles. The infimum of its spectrum is $(\hbar/2)\sum_j\mu_j$, which proves (\ref{eq:offset}). The equivalences follow from the normal form and from Theorem~\ref{thm:reconstruction}.
\end{proof}

\begin{corollary}[Quadratic masters of constraints]\label{cor:dirac}
For the master (\ref{eq:quadmaster}) with independent linear constraints $C_i$, $\inf\spec\Op(m)=\frac\hbar2\Tr_+(K\Dmat)$, with $\Dmat$ the constant Dirac matrix. For a single second-class pair with $K=\mathbf 1$ this is $\frac\hbar2|\{C_1,C_2\}|$, which saturates the central-term bound $\ev{\Chat_1^2}+\ev{\Chat_2^2}\ge\hbar|\{C_1,C_2\}|$ of \cite{Corichi2008}. When all the constraints are second class the infimum is attained; the minimizing states are the Gupta--Bleuler states of \cite{Corichi2008}, and different weights select vacua related by relative squeezing \cite{CorichiCortez}.
\end{corollary}

\begin{proof}
Combine Proposition~\ref{prop:offset} with (\ref{eq:trKC}). The central-term bound follows from $\ev{\Chat_1^2}+\ev{\Chat_2^2}\ge2(\ev{\Chat_1^2}\ev{\Chat_2^2})^{1/2}\ge|\ev{[\Chat_1,\Chat_2]}|$. In the normal form (\ref{eq:williamson}) the ground state is annihilated by the annihilation operators of the rotation blocks, which is the Gupta--Bleuler condition. For one pair, the ground state of $\frac12(k_1C_1^2+k_2C_2^2)$ is the vacuum of the compatible metric proportional to ${\rm diag}(k_1,k_2)$ in the chart $(C_1,C_2)$.
\end{proof}

For general master operators the spectrum is not explicitly computable, but its bottom can be bounded from above by trial states. In the following, a \emph{master operator with symbol $m$} is a self-adjoint operator $\Mhat_\hbar$ whose quadratic form on the Schwartz space $\mathcal S(\R^n)$ is that of $\Op(m+\hbar^2m_2)$, with $m_2$ smooth and of polynomial growth with all its derivatives. Master operators built as $\frac12K^{ij}\Chat_i\Chat_j$ from Weyl-quantized constraints with constant symmetric $K$ are of this form: the term of order $\hbar$ in the Weyl symbol, $\frac{i\hbar}4K^{ij}\{C_i,C_j\}$, cancels by symmetry.

\begin{proposition}[General masters]\label{prop:general}
Let $m\ge0$ be smooth, with all derivatives of polynomial growth, and $\Sigma=m^{-1}(0)$. Let $\Mhat_\hbar$ be a master operator with symbol $m$. Then, for $0<\hbar\le1$:
\begin{enumerate}[(a)]
\item for every $z\in\Sigma$, there is a constant $C_z$ such that
\begin{equation}\label{eq:generalA}
\inf\spec\Mhat_\hbar\le\frac\hbar2\Tr_+L_z+C_z\,\hbar^{4/3};
\end{equation}
\item if $\Sigma$ is a submanifold that is coisotropic on a nonempty open subset, in particular if it is a regular first-class constraint surface, there is a constant $C$ such that
\begin{equation}\label{eq:generalB}
\inf\spec\Mhat_\hbar\le C\hbar^2 .
\end{equation}
\end{enumerate}
\end{proposition}

The proof, in Appendix~\ref{app:general}, uses two families of trial states. For (a) it is a Gaussian packet at $z$, in the ground state of the rotation blocks of the normal form of $\Hess_zm$ and squeezed at the scale $\hbar^{1/2\pm1/6}$ in the nilpotent blocks. For (b) it is a Lagrangian (WKB) state concentrated on a Lagrangian submanifold contained in $\Sigma$, which exists exactly where $\Sigma$ is coisotropic; the term of order $\hbar$ in its energy vanishes because the amplitude can be chosen real. No harmonic approximation is assumed.

\begin{remark}[Lower bounds and the scope of the spectral test]\label{rem:lower}
A matching lower bound, $\inf\spec\Mhat_\hbar\ge\frac\hbar2\inf_{z\in\Sigma}\Tr_+L_z-o(\hbar)$, is the semiclassical content of Melin's inequality for nonnegative symbols that vanish exactly to second order on $\Sigma$ \cite{Melin,Hormander85}; it requires uniformity assumptions at infinity that we do not discuss. For magnetic Schr\"odinger operators the resulting asymptotics is established in \cite{HelfferMohamed}; see \cite{Raymond} for a review. Proposition~\ref{prop:general} and this lower bound together give the harmonic approximation $\inf\spec\Mhat_\hbar=\frac\hbar2\inf_\Sigma\Tr_+L+o(\hbar)$. Since this involves a \emph{global infimum}, the order of the offset tests whether $\Tr_+L_z$ vanishes somewhere on $\Sigma$, not whether $\Sigma$ is first class. The next example shows that the difference is real.
\end{remark}

\begin{example}[Magnetic master constraints]\label{ex:magnetic}
On $T^*\R^2$ take $C_i=\xi_i-A_i(x)$, with $\{C_1,C_2\}=B=\partial_1A_2-\partial_2A_1$, and $m=\frac12(C_1^2+C_2^2)$. Then $\Op(m)=\frac12(\hbar D-A)^2$ exactly, the Hamiltonian of a charged particle in the magnetic field $B$, and $\Sigma=\{\xi=A(x)\}$ is two-dimensional. By (\ref{eq:LDirac}), $L_z$ acts on its image as the rotation with frequency $|B(x)|$, so $\Tr_+L_z=|B(x)|$. Where $B\neq0$ the two constraints are second class; where $B=0$, $T_z\Sigma$ is Lagrangian.
\begin{enumerate}[(i)]
\item For constant $B$, the offset $\hbar|B|/2$ is the lowest Landau level. Subtracting it and restricting to the lowest level is the standard reduction of this second-class system to the quantization of its reduced, Dirac-bracket phase space \cite{DJT}, which parallels the subtraction of the master offset in \cite{DT2,DT3}.
\item For $B\ge b_0>0$ the offset is $\frac\hbar2\inf|B|+o(\hbar)$ \cite{HelfferMohamed}. For $A=(0,x_1+x_1^3/3)$, i.e.\ $B=1+x_1^2$, the ratio $\inf\spec\Op(m)/\hbar$ is $0.5234$, $0.5073$, $0.5025$ and $0.5007$ for $\hbar=0.1$, $0.03$, $0.01$ and $0.003$.
\item For $A=(0,x_1^2/2)$, i.e.\ $B=x_1$, the operator separates, and scaling gives $\inf\spec\Op(m)=\frac12E_*\hbar^{4/3}$ exactly, where $E_*\simeq0.5698$ is the minimum over $\kappa$ of the ground-state energy of $-\partial_s^2+(\kappa-s^2/2)^2$. The same leading behaviour holds whenever $B$ vanishes nondegenerately along a closed curve with $|\nabla B|=1$ there \cite{Montgomery}. The offset is $o(\hbar)$, although $\Sigma$ is symplectic away from the zero locus of $B$ and coisotropic on no open set. The exponent is that of (\ref{eq:generalA}), which therefore cannot be improved in general. For comparison, a Gaussian trial state centred on the zero locus, of the type used in Appendix~\ref{app:general} with optimized widths, gives $\frac12\times0.6814\,\hbar^{4/3}$.
\end{enumerate}
\end{example}

We can now state precisely what the order of the offset says about a master constraint. If the zero set is a regular first-class surface, Proposition~\ref{prop:general}(b) gives an offset of order $\hbar^2$ at most; for a positive master operator the offset lies in $[0,C\hbar^2]$, whatever the degree of the master function. This establishes, for general masters, the part of the observation of Dittrich and Thiemann that concerns first-class constraints: their offsets are at most of order $\hbar^2$, as expected of normal-ordering effects \cite{DT3}. If instead every point of $\Sigma$ carries second-class directions, uniformly in the sense $\inf_\Sigma\Tr_+L>0$, Melin's lower bound gives an offset of order $\hbar$, as for the second-class systems of \cite{DT2}; subtracting it then selects Gupta--Bleuler states, as in Corollary~\ref{cor:dirac} and Example~\ref{ex:magnetic}(i), and does not turn second-class constraints into first-class ones. In intermediate situations, the order of the offset is set by the point of $\Sigma$ that is least second class. The examples below conform to these statements. The second-class pair gives $\inf\spec\Op((q^2+p^2)/2)=\hbar/2$; the fixed-energy oscillator gives an offset of at most $\hbar^2\nu^2/8$ (Section~\ref{sec:oscillator}); and the SL(2,$\R$) model gives exactly $\hbar^2/2$ (Section~\ref{sec:bottom}).

\section{Semiclassical data and the fluctuation diagnostic}\label{sec:semiclassical}

The spectral test of Section~\ref{sec:offset} needs only the family of operators $\Mhat_\hbar$. The other two quantum signatures of the gauge structure need more. The static offset of the master expectation requires a family of semiclassical states, and the dynamical shear requires in addition the master unitary. We now formulate these requirements.

\subsection{What a semiclassical identification supplies}

An abstract physical Hilbert space does not determine a unique unreduced classical gauge theory. To ask for one, one must specify semiclassical observables and a family of states identifying their macroscopic values with points of a manifold $\Gamma$. This is additional physical input; as emphasized in \cite{ABC}, semiclassicality is always relative to a chosen set of observables and tolerances.

Let $z\mapsto[\Psi_\hbar(z)]$ be a smooth family of kinematical rays $\Gamma\to\Ps_\hbar$, and define
\begin{equation}\label{eq:pullbacks}
\omega_\hbar=\Psi_\hbar^*\Omega_\hbar,\qquad m_\hbar=\Psi_\hbar^*\ev{\Mhat_\hbar}.
\end{equation}
We require a nondegenerate limiting symplectic form $\omega$, convergence $\omega_\hbar\to\omega$ in $C^1$ and $m_\hbar\to m_0$ in $C^2$, locally uniformly on compact sets. These conditions control the HVFs and their linearizations. The limiting zero set $\Sigma=m_0^{-1}(0)$ must independently satisfy the Morse--Bott hypothesis.

There is no reason for $m_\hbar$ to vanish at a classical constrained point. The expectation of a square includes the variance, so a coherent packet with vanishing constraint mean has a master expectation of order $\hbar$. This is the dispersion term of the decomposition $\ev{\Chat^2}=\ev{\Chat}^2+(\Delta\Chat)^2$; for a single constraint and isotropic coherent states centred on $\Sigma$ it equals $\frac\hbar2|X_C|^2+O(\hbar^2)$ \cite{Corichi2008}. The critical manifold emerges only in the limit, and at nonzero $\hbar$ the packets probing $\Sigma$ are kinematical states, not physical solutions.

Moreover, the HVF of $m_\hbar$ with respect to $\omega_\hbar$ need not be the restriction of the ambient quantum HVF of $\ev{\Mhat_\hbar}$, since the quantum flow can leave the chosen family. Correct expectation values alone do not establish the propagation of actual quantum states. We check propagation directly: by exact formulas for the oscillator, and by a norm estimate for the non-Abelian model.

\subsection{Gauge directions in the leading fluctuations}

Suppose a wave packet centred at $z\in\Sigma$ has a controlled Gaussian approximation under the master evolution, and let $V_0>0$ be its initial covariance divided by $\hbar$, in fixed Darboux coordinates. At a fixed centre, the leading covariance is transported by the tangent map. When $L_z^2=0$,
\begin{equation}\label{eq:Vt}
V_z(t)=(1+tL_z)\,V_0\,(1+tL_z)^{\rm T},
\end{equation}
and
\begin{equation}\label{eq:Vddot}
\ddot V_z(t)=2L_zV_0L_z^{\rm T},\qquad \Im\ddot V_z=\Im L_z=D_z .
\end{equation}
The image equality uses only $V_0>0$ and is independent of the initial squeezing (Appendix~\ref{app:probes}). The covariance is a symmetric bivector, so its image is a tangent subspace and transforms as such under a change of Darboux frame; no Euclidean metric on $\Gamma$ is involved. If $\Sigma$ has second-class directions at $z$, the corresponding part of the covariance rotates with the frequencies $\mu_j$ instead of growing.

Equation (\ref{eq:Vt}) is the transport of a Gaussian by the linearized classical flow, and a classical Liouville ensemble of width $\sqrt\hbar$ would obey the same law. The quantum input is the theorem that wave packets follow the linearized classical dynamics with a controlled error \cite{Heller,Hagedorn,Littlejohn,CombescureRobert}, which here must be established for the master unitary. Granting it, the packet centre may stay fixed while the leading fluctuations respond along the gauge directions. These directions are not degenerate directions of the quantum symplectic form (Proposition~\ref{prop:quantum}); they emerge from the semiclassical response of kinematical states. The master unitary is not a physical evolution, and $t$ is a parameter of the kinematical construction. The diagnostic is a statement about the dynamics generated by $\Mhat$, computable within the theory, not a laboratory protocol.

The statement can be formulated without assuming convergence of unbounded second moments of the exact evolved state. Let $\widehat Z$ be the vector of canonical operators and use the bounded unitaries
\begin{equation}\label{eq:probes}
\widehat W_{z,\hbar}(\xi)=\exp\Bigl(\frac{i}{\sqrt\hbar}\,\xi^{\rm T}(\widehat Z-z)\Bigr),\qquad\xi\in\R^{2n}.
\end{equation}
If the quantum state approaches the leading Gaussian in norm, their expectations approach
\begin{equation}\label{eq:charfn}
\chi_z(t,\xi)=\exp\Bigl(-\frac12\,\xi^{\rm T}V_z(t)\,\xi\Bigr).
\end{equation}
The covariance in (\ref{eq:Vddot}) is defined by this limiting characteristic function: one first takes the limit of bounded expectations and then differentiates the limit, so no interchange of limits is required. The probes are bounded although they resolve the fluctuation scale $\sqrt\hbar$. Section~\ref{sec:sl2r} proves this limit, with a uniform error, for the non-Abelian example, and Section~\ref{sec:oscillator} computes the exact covariance in the Abelian one.

\begin{remark}[Two moments of one matrix]\label{rem:moments}
For the quadratic master (\ref{eq:quadmaster}) and isotropic coherent packets, $V_0=\frac12$ in a Darboux frame in which the coherent-state metric is Euclidean. With $G_{ij}$ the Gram matrix of Proposition~\ref{prop:kahler}, one has on $\Sigma$
\begin{equation}\label{eq:moments}
m_\hbar-m_0=\frac\hbar4\Tr(KG)+O(\hbar^2),\qquad \Tr\ddot V_z=\Tr\bigl((KG)^2\bigr).
\end{equation}
The first follows from the Gaussian smoothing $e^{(\hbar/4)\Delta}$ of the Weyl symbol $C_iC_j+O(\hbar^2)$ of $\frac12(\Chat_i\Chat_j+\Chat_j\Chat_i)$, whose $O(\hbar)$ term on $\Sigma$ is $\frac\hbar2\,g(\nabla C_i,\nabla C_j)$; it is the orbit-speed term of \cite{Corichi2008}. The second follows from $\ddot V_z=L_zL_z^{\rm T}$ and $L_z=\Pi A^{\rm T}KA$, where the rows of $A$ are the gradients of the $C_i$. Thus the static $O(\hbar)$ offset of the master expectation and the rate at which packets are sheared are the first and second moments of the same matrix $KG$. In the oscillator of Section~\ref{sec:oscillator} they are $\hbar\nu E/2$ and $4\nu^4S^2$; at the SL(2,$\R$) point (\ref{eq:regularpoint}) they are $3\hbar r^2$ and $48r^4$. For a compact gauge group acting linearly, a third quantity built from the same Gram matrix, the volume of the gauge orbit, sets the probability that a coherent packet centred on $\Sigma$ is physical: up to $O(\hbar)$ it is $(4\pi\hbar)^{k/2}$ divided by that volume, with $k$ the orbit dimension \cite{CorichiDirac}. For the oscillator see Section~\ref{sec:oscphys}.
\end{remark}

\subsection{Relation to near-constraint evolution and to effective constraints}

Gauge motions can also be recovered from a master evolution near the constraint surface. In uniform discretizations one approaches the surface with $C_i=c_i$ and rescales the evolution parameter; for a quadratic master, $\eps^{-1}X_m$ tends to a combination of constraint fields with multipliers given by the limiting $c_i$ \cite{CDGP,GambiniOlmedo}. The linearization gives the same structure,
\begin{equation}\label{eq:nearconstraint}
\eps^{-1}X_m(z+\eps\,\delta z)\longrightarrow L_z\,\delta z ,
\end{equation}
uniformly for bounded $\delta z$. We do not claim this near-constraint recovery as new. What we add is its intrinsic content: which subspace is obtained, when it is tangent and integrable, how many constraints of each class it reflects, how it appears in the spectrum, and how it is recorded by quantum packet fluctuations.

A related semiclassical framework is the theory of effective constraints \cite{BSST,BT}. There, expectation values and moments of the constraint operators are imposed as constraints on a quantum phase space of moments, and the gauge flows they generate act on fluctuations as well as on expectation values. The present construction differs in two respects. A single positive master operator is used, whose moments contain no first-order gauge generator on $\Sigma$. And the gauge directions are read off from the response of the leading covariance under the exact master unitary, not from flows generated by constraint moments. A comparison of the two moment pictures, in which the fluctuation terms of the master symbols below would appear as effective-constraint corrections, is left for future work.

\section{Two oscillators with fixed total energy}\label{sec:oscillator}

The first example keeps three spaces in view: the kinematical phase space, the presymplectic constraint surface, and a semiclassical family in the physical quantum space. It also gives the master unitary exactly, so that the reconstruction can be checked without a general propagation theorem.

\subsection{Classical surface, null circles and quotient}

For two oscillators of equal frequency $\nu$, with unit mass, use the action-scaled coordinates
\begin{equation}\label{eq:actionscaled}
z_j=\frac1{\sqrt2}\Bigl(\sqrt\nu\,q_j+\frac{ip_j}{\sqrt\nu}\Bigr),\qquad S=|z_1|^2+|z_2|^2,\qquad \omega_{\rm kin}=i\sum_j{\rm d}z_j\wedge{\rm d}\bar z_j .
\end{equation}
Let $E>0$, $C=\nu S-E$ and $m_0=C^2/2$. The zero set is the three-sphere $S=I:=E/\nu$. For a variation $\delta z=(\delta z_1,\delta z_2)$,
\begin{equation}\label{eq:oscdC}
{\rm d}C_z(\delta z)=2\nu\,{\rm Re}\sum_j\bar z_j\delta z_j,\qquad X_C(z)=-i\nu z ,
\end{equation}
so $X_C$ rotates both coordinates by a common phase, and on the surface
\begin{equation}\label{eq:oscL}
L_z(\delta z)={\rm d}C_z(\delta z)\,X_C(z),\qquad L_z^2=0 .
\end{equation}
The image of $L$ is the common-phase direction: a radial variation produces a phase displacement, and variations tangent to the sphere are annihilated. In the language of Proposition~\ref{prop:kahler}, the radial direction is $JX_C$, the imaginary gauge direction, and $KG=|X_C|^2=2\nu E$. The characteristic leaves are the Hopf circles and the quotient is $\CP^1$. With $w=z_2/z_1$,
\begin{equation}\label{eq:oscred}
\omega_{\rm red}=I\,\omega_{\rm FS},\qquad \omega_{\rm FS}=i\partial\bar\partial\log(1+|w|^2),\qquad \int_{\CP^1}\omega_{\rm FS}=2\pi ,
\end{equation}
since the section $z=\sqrt I(1,w)/\sqrt{1+|w|^2}$ pulls the restricted canonical form back to (\ref{eq:oscred}). The reduced area is $2\pi E/\nu$. Physically, the constraint fixes the total action and the common phase is redundant, while the relative amplitude and phase remain as coordinates on the reduced sphere.

\subsection{Exact coherent symbol and quantum master evolution}

The kinematical Hilbert space is the two-mode Fock space. We retain the zero-point term in the energy operator, which makes the relation between the physical energy shell and its induced symplectic form explicit. With $\hat n=\widehat N_1+\widehat N_2$, let
\begin{equation}\label{eq:oscC}
\Chat_\hbar=\hbar\nu(\hat n+1)-E,\qquad \Mhat_\hbar=\tfrac12\Chat_\hbar^2 .
\end{equation}
The Weyl symbol of $\Chat_\hbar$ is exactly $\nu S-E$. The standard coherent state with labels $\alpha_j=z_j/\sqrt\hbar$ has a Poisson-distributed total number with mean and variance $R=S/\hbar$. Therefore (Appendix~\ref{app:oscillator})
\begin{equation}\label{eq:oscsymbol}
m_\hbar(z)=\tfrac12(\nu S+\hbar\nu-E)^2+\tfrac12\hbar\nu^2S .
\end{equation}
The symplectic pullback of the coherent family is exactly $\omega_{\rm kin}$, and (\ref{eq:oscsymbol}) converges to $m_0$ with all derivatives on compact sets. A seed on the classical surface is not an exact solution: on $S=E/\nu$,
\begin{equation}\label{eq:oscoffset}
m_\hbar|_\Sigma=\tfrac12\hbar\nu E+\tfrac12\hbar^2\nu^2 ,
\end{equation}
whose leading term is half the variance of $\Chat_\hbar$, in agreement with (\ref{eq:moments}). A sharpening kinematical packet thus probes the classical surface although no member of the family satisfies the constraint. The spectrum of $\Mhat_\hbar$ is $\{\frac12(\hbar\nu(n+1)-E)^2\}_{n\ge0}$, so for $\hbar\nu\le E$ its offset is at most $\hbar^2\nu^2/8$, of order $\hbar^2$ as Proposition~\ref{prop:general}(b) requires for the coisotropic three-sphere.

The master evolution $U_\hbar(t)=\exp(-it\Mhat_\hbar/\hbar)$ is generally not a coherent-state motion. Nevertheless the centre $Z_{j,\hbar}(t)=\sqrt\hbar\,\ev{\hat a_j}_t$ is exactly
\begin{equation}\label{eq:osccentre}
Z_{j,\hbar}(t)=z_j\,e^{it\nu E-\frac32it\hbar\nu^2}\exp\Bigl[\frac S\hbar\bigl(e^{-it\hbar\nu^2}-1\bigr)\Bigr].
\end{equation}
To derive it, let $\mathcal E(n)=[\hbar\nu(n+1)-E]^2/2$. The expectation of $\hat a_j$ involves $\mathcal E(k+1)-\mathcal E(k)$ with $k$ Poisson distributed with mean $R$, and $[\mathcal E(k+1)-\mathcal E(k)]/\hbar=\hbar\nu^2(k+\frac32)-\nu E$. Summing the generating function gives (\ref{eq:osccentre}). For bounded $t$ and centres in a compact set,
\begin{equation}\label{eq:osclimit}
Z_{j,\hbar}(t)\longrightarrow z_j\,e^{-it\nu(\nu S-E)}
\end{equation}
in every $C^k$ norm with respect to the real centre coordinates: after the leading term $-it\nu^2S$ of the exponent cancels, the remainder and all its derivatives are $O(\hbar)$. The limit is the classical master flow, which fixes the centre on $\Sigma$ and has derivative $1+tL$ there.

The packets remain localized. In action-scaled real quadratures $(Q_j,P_j)$, with $z_j=(Q_j+iP_j)/\sqrt2$, conservation of $\ev{\hat n}$ and (\ref{eq:osccentre}) give
\begin{equation}\label{eq:oscspread}
\sum_{j=1}^2\bigl[(\Delta Q_j)^2+(\Delta P_j)^2\bigr]_t=2\hbar+2S\Bigl[1-e^{\frac{2S}\hbar(\cos(t\hbar\nu^2)-1)}\Bigr]=2\hbar+2S^2\hbar\nu^4t^2+O(\hbar^2).
\end{equation}
At fixed $t$ the total spread is therefore $O(\hbar)$, and the convergence of centres describes localized packets, not only a formal limit of functions. The unitary creates phase correlations between number components, so the packet need not remain a standard coherent state: expectation-value evolution, localization and membership in a preferred family are distinct questions. The parameter $t$ labels the master unitary and has units of $\hbar/E^2$; it is not physical oscillator time. All limits above hold on bounded intervals of $t$. The exact formula also displays collapse and revival at times of order $1/(\hbar\nu^2)$, outside the fixed-time correspondence considered here.

\subsection{Exact covariance and the gauge shear}

The coefficient of $t^2$ in (\ref{eq:oscspread}) is not accidental. For $V_0=\frac12$ and $L_z$ given by (\ref{eq:oscL}), $\hbar\Tr(L_zV_0L_z^{\rm T})=\frac\hbar2|X_C|^2|\nabla C|^2=2\hbar\nu^4S^2$, which is exactly that coefficient. In fact the whole covariance can be computed. Besides the first moments (\ref{eq:osccentre}) one needs the pair moment
\begin{equation}\label{eq:pairmoment}
\ev{\hat a_j\hat a_k}_t=\alpha_j\alpha_k\exp\Bigl[2it\nu E-4it\hbar\nu^2+\frac S\hbar\bigl(e^{-2it\hbar\nu^2}-1\bigr)\Bigr],
\end{equation}
obtained in the same way from $\mathcal E(k+2)-\mathcal E(k)$, while $\ev{\hat a_j^\dagger\hat a_k}=\bar\alpha_j\alpha_k$ is conserved because $\Mhat_\hbar$ commutes with every number-preserving bilinear.

\begin{proposition}\label{prop:osc-cov}
Let $V_{z,\hbar}(t)$ be the symmetrized covariance matrix of the four quadratures $Q_j,P_j$ in the state $U_\hbar(t)|z,\hbar\rangle$, divided by $\hbar$. For $z\in\Sigma$ and bounded $t$,
\begin{equation}\label{eq:osccov}
V_{z,\hbar}(t)=\tfrac12\,(1+tL_z)(1+tL_z)^{\rm T}+O(\hbar),
\end{equation}
uniformly for $z$ in compact subsets of $\Sigma$. The second derivative $\ddot V_z=L_zL_z^{\rm T}$ has rank one, and its image is spanned by $X_C(z)=-i\nu z$, the Hopf direction.
\end{proposition}

\begin{proof}
Write $\varphi=t\hbar\nu^2$ and $R=S/\hbar$. Expanding the exponents of (\ref{eq:osccentre}) and (\ref{eq:pairmoment}) to first order in $\hbar$ at fixed $t$ gives (Appendix~\ref{app:oscillator})
\[
{\rm Cov}(\hat a_j,\hat a_k)=-\frac{z_jz_k}{\hbar}\bigl(it\hbar\nu^2+t^2\hbar\nu^4S\bigr)e^{2it\nu(E-\nu S)}+O(\hbar),\qquad {\rm Cov}(\hat a_j^\dagger,\hat a_k)=\frac{\bar z_jz_k}{\hbar}\,t^2\hbar\nu^4S+O(\hbar).
\]
On $\Sigma$ the phase equals one. Converting to quadratures with $\hat a_j=(\widehat Q_j+i\widehat P_j)/\sqrt{2\hbar}$, the terms linear in $t$ reproduce $\frac t2(L_z+L_z^{\rm T})$ and the quadratic ones $\frac{t^2}2L_zL_z^{\rm T}$, where $L_z=X_C\,\nabla C^{\rm T}$ with $\nabla C=\nu(Q_1,P_1,Q_2,P_2)$ and $X_C=\Pi\nabla C$. The rank and image statements follow from $L_zL_z^{\rm T}=|\nabla C|^2X_CX_C^{\rm T}$.
\end{proof}

\begin{figure}[t]
\centering
\includegraphics[width=\textwidth]{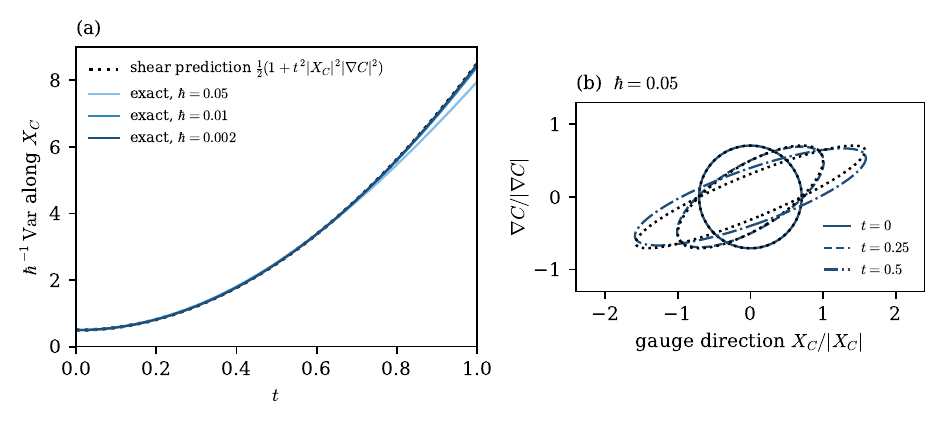}
\caption{Exact master evolution of a coherent packet centred at $z=(1,i)$ on the constraint surface $S=E/\nu$ of two oscillators ($\nu=1$, $E=2$). (a) Variance along the gauge direction $X_C$, divided by $\hbar$, from the exact moments (\ref{eq:osccentre}) and (\ref{eq:pairmoment}), compared with the shear prediction $\frac12(1+t^2|X_C|^2|\nabla C|^2)$ of (\ref{eq:osccov}). Deviations are $O(\hbar)$ at fixed $t$. (b) Covariance ellipses in the plane spanned by the gauge direction and the transverse direction $\nabla C=JX_C$, for $\hbar=0.05$ (solid, dashed, dash-dotted), with the shear prediction dotted. Unlike Figure~\ref{fig:shear}, these are exact quantum moments.}
\label{fig:osc}
\end{figure}

The oscillator therefore realizes the fluctuation diagnostic of Section~\ref{sec:semiclassical} exactly, without a propagation estimate. Figure~\ref{fig:osc} shows the exact covariance, computed from (\ref{eq:osccentre}) and (\ref{eq:pairmoment}), against (\ref{eq:osccov}). For $\nu=1$, $E=2$, the centre $z=(1,i)$ and $t=\frac12$, the maximal entry of the difference is $1.5\times10^{-2}$, $1.5\times10^{-3}$ and $1.5\times10^{-4}$ for $\hbar=10^{-2}$, $10^{-3}$ and $10^{-4}$; the coefficient of $\hbar$ depends on the centre. The packet stays centred at a fixed classical point, its transverse width stays fixed, and its width along the Hopf circle grows linearly in $t$, at the rate fixed by the transverse width and by $KG$.

\subsection{The reduced physical symplectic limit}\label{sec:oscphys}

To compare with the physical theory, the constraint must be imposed. For fixed $E$ a nontrivial normalizable kernel exists only if $E=\hbar\nu(N+1)$ for some integer $N\ge0$. We therefore use a sequence of quantum theories with the same classical energy and $\hbar_N=E/[\nu(N+1)]$, so that exact physical vectors occupy the total-number-$N$ subspace, whose ray space is $\CP^N$. Projecting a coherent seed and normalizing gives
\begin{equation}\label{eq:wN}
|w;N\rangle=\frac{(\hat a_1^\dagger+w\hat a_2^\dagger)^N}{\sqrt{N!}\,(1+|w|^2)^{N/2}}\,|0\rangle .
\end{equation}
For $N\ge1$ these rays form a Veronese copy of $\CP^1$ in $\CP^N$, the familiar two-mode fixed-number, or SU(2) spin, coherent states \cite{Radcliffe,ACGT,Perelomov}. The projection construction belongs to the group-averaged semiclassical framework of \cite{ABC}, and its geometry, including the bundle of coherent seeds over this Veronese curve, is analysed in \cite{CorichiGA}. The holomorphic unnormalized vector has squared norm $(1+|w|^2)^N$, so the induced symplectic form is
\begin{equation}\label{eq:physlimit}
\Omega_{{\rm phy},N}=\hbar_NN\,\omega_{\rm FS}\longrightarrow(E/\nu)\,\omega_{\rm FS}=\omega_{\rm red} .
\end{equation}
The comparison uses the selected coherent family, not every ray of $\CP^N$. The deficit $I-\hbar_NN=\hbar_N$ is the zero-point correction. This agreement of the quantization of the reduced sphere with the reduction of the quantum theory is an instance of the Guillemin--Sternberg principle \cite{GuilleminSternberg}. On this family the number-preserving Schwinger operators, which commute with the constraint, have mean vector of length $\hbar_NN/2\to I/2$, the reduced-sphere radius, and transverse variances $\hbar_N^2N/4=O(\hbar_N)$ \cite[section~4]{CorichiGA}. The family is therefore semiclassical for the reduced observables.

The kinematical shear of the previous subsection and the physical symplectic limit concern different stages of the comparison, and both are realized in the same example. Seeds related by a common phase of $z$ project to the same physical ray, but so do seeds differing by a rescaling of $z$, so the averaging fibre is larger than a gauge orbit; within the coherent family it is exactly the complexified gauge orbit \cite{CorichiGA}. The projection nevertheless distinguishes the points of such an orbit. With $P_N$ the projector onto total number $N$, the probability that a coherent seed is physical is
\begin{equation}\label{eq:oscweight}
\|P_N|z,\hbar_N\rangle\|^2=e^{-R}\,\frac{R^N}{N!},\qquad R=S/\hbar_N .
\end{equation}
Along each complexified orbit $R$ takes every positive value, and (\ref{eq:oscweight}) is maximal at $R=N$, where the constraint expectation $\hbar_N\nu(R-N)$ vanishes: the projection singles out the constraint surface, shifted by the zero-point term, as the place where seeds are most physical. The maximal value is $e^{-N}N^N/N!=(2\pi N)^{-1/2}\bigl(1+O(1/N)\bigr)$. On the circle $S=\hbar_NN$ the Hopf circle has length $\ell=2\pi|X_C|/\nu$ in the coherent-state metric, and $(4\pi\hbar_N)^{1/2}/\ell=(2\pi N)^{-1/2}$ exactly. The probability that a seed on the constraint surface is physical is thus one over the number of Planck cells needed to cover its gauge orbit, an instance of a law established in \cite{CorichiDirac} for linear actions of compact groups and for heat-kernel packets. We return to this point in Section~\ref{sec:realimag}.

\section{The non-Abelian \texorpdfstring{SL(2,$\R$)}{SL(2,R)} model}\label{sec:sl2r}

The second example tests the construction for a non-Abelian algebra with several gauge directions and a noncompact gauge group. It also shows that two quantum prescriptions can share the leading classical gauge geometry and still require different physical-sector input. The model is finite-dimensional with constant structure coefficients, so it does not reproduce the structure functions of gravity.

\subsection{Constraints, HVFs and the regular surface}

On $\Gamma=T^*\R^4$, with coordinates $(u,v,p,\pi)$ consisting of two-component real vectors, take
\begin{equation}\label{eq:sl2omega}
\omega_{\rm kin}=\sum_{a=1}^2\bigl({\rm d}u_a\wedge{\rm d}p_a+{\rm d}v_a\wedge{\rm d}\pi_a\bigr).
\end{equation}
The model of \cite{MRT} has the constraints
\begin{equation}\label{eq:sl2constraints}
H_1=\tfrac12(p^2-v^2),\qquad H_2=\tfrac12(\pi^2-u^2),\qquad D=u\cdot p-v\cdot\pi ,
\end{equation}
with
\begin{equation}\label{eq:sl2algebra}
\{H_1,H_2\}=D,\qquad \{H_1,D\}=-2H_1,\qquad \{H_2,D\}=2H_2 .
\end{equation}
In the order $(u,v,p,\pi)$, their HVFs are
\begin{equation}\label{eq:sl2HVF}
X_{H_1}=(p,0,0,v),\qquad X_{H_2}=(0,\pi,u,0),\qquad X_D=(u,-v,-p,\pi).
\end{equation}
The finite gauge transformations act by the same SL(2,$\R$) matrix on each column $(u_a,p_a)^{\rm T}$ and $(\pi_a,v_a)^{\rm T}$.

On the regular part of $\Sigma=\{H_1=H_2=D=0\}$, the surface has dimension five, the characteristic distribution dimension three, and the reduced space dimension two. An explicit regular point is
\begin{equation}\label{eq:regularpoint}
u=v=(r,0),\qquad p=(0,r),\qquad \pi=(0,-r),\qquad r>0 .
\end{equation}
Let $A$ be the $3\times8$ matrix whose rows are the gradients of $(H_1,H_2,D)$, and put
\begin{equation}\label{eq:sl2L}
K={\rm diag}(2,2,1),\qquad L=\Pi A^{\rm T}KA .
\end{equation}
At (\ref{eq:regularpoint}) the three gradients are mutually orthogonal with squared lengths $2r^2$, $2r^2$, $4r^2$, and the algebra gives $A\Pi A^{\rm T}=0$ on $\Sigma$, i.e.\ a vanishing Dirac matrix. Hence $L^2=0$, and the three nonzero singular values of $L$ all equal $4r^2$ (Appendix~\ref{app:probes}).

The master function
\begin{equation}\label{eq:sl2master}
m_0=H_1^2+H_2^2+\tfrac12D^2
\end{equation}
is one half of the master function $D^2+(H_1+H_2)^2+(H_1-H_2)^2$ used in \cite[equation~(3.17)]{DT3}. Its linearized HVF on the regular surface is
\begin{equation}\label{eq:sl2Lz}
L_z(\delta z)=2\,{\rm d}H_1(\delta z)X_{H_1}+2\,{\rm d}H_2(\delta z)X_{H_2}+{\rm d}D(\delta z)X_D ,
\end{equation}
and Theorem~\ref{thm:reconstruction} gives $\rank L=3$, $\ker L=T\Sigma$, $\Im L=D$ and $L^2=0$: three first-class and no second-class constraints. A single master function thus reconstructs the three-dimensional gauge distribution. In the frame $X_{H_1},X_{H_2},X_D$ of this distribution one recovers (\ref{eq:sl2algebra}), with the sign of (\ref{eq:bracket}); other frames of the same distribution have other structure functions (Section~\ref{sec:algebra}).

One regular component is parametrized by the section
\begin{equation}\label{eq:sl2section}
u=(\sqrt\rho,0),\quad p=(0,\sqrt\rho),\quad v=\sqrt\rho\,(\cos\phi,\sin\phi),\quad \pi=\sqrt\rho\,(\sin\phi,-\cos\phi),\qquad \rho>0 .
\end{equation}
The canonical one-form $\theta=p\cdot{\rm d}u+\pi\cdot{\rm d}v$ pulls back to $-\rho\,{\rm d}\phi$; since $\omega_{\rm kin}=-{\rm d}\theta$, the reduced symplectic form on this component is ${\rm d}\rho\wedge{\rm d}\phi$. On this section both angular momenta $O_{12}=u_1p_2-u_2p_1$ and $O_{34}=\pi_1v_2-\pi_2v_1$ equal $\rho>0$; the other components require other sign choices. The global quotient contains an exceptional non-Hausdorff set \cite{LoukoRovelli}, so no single smooth quotient describes the whole model.

The Morse--Bott hypothesis fails on the singular part of $\Sigma$, which is larger than the origin. On $\{p=v=0,\ |u|=|\pi|\}\subset\Sigma$ one has ${\rm d}H_1=0$, and on $\{u=\pi=0,\ |p|=|v|\}$ one has ${\rm d}H_2=0$; on both the gradients have rank two, and at the origin all three vanish. Theorem~\ref{thm:reconstruction} applies on the complement of these strata.

\subsection{Quantum constraints, coherent expectations and Weyl symbol}

On $L^2(\R^4)$ take the Weyl-ordered constraints
\begin{equation}\label{eq:sl2quantum}
\widehat H_1=\tfrac12\bigl(-\hbar^2\Delta_u-v^2\bigr),\qquad \widehat H_2=\tfrac12\bigl(-\hbar^2\Delta_v-u^2\bigr),\qquad \widehat D=-i\hbar\bigl(u\cdot\nabla_u-v\cdot\nabla_v\bigr).
\end{equation}
The symmetric-ordering constants in $\widehat D$ cancel because the two sectors have equal dimension. On Schwartz vectors the commutators reproduce (\ref{eq:sl2algebra}) multiplied by $i\hbar$; quadratic Weyl operators have no higher Moyal corrections to their commutators.

Let $|z,\hbar\rangle$ be the standard four-mode coherent packet with centre $z=(u,v,p,\pi)$ and covariance $\hbar\mathbf 1_8/2$. All three constraint means equal their classical values. A direct Gaussian computation (Appendix~\ref{app:symbols}) gives, with $|z|^2=u^2+v^2+p^2+\pi^2$,
\begin{align}
(\Delta H_1)^2&=\frac\hbar2(p^2+v^2)+\frac{\hbar^2}2,\label{eq:varH1}\\
(\Delta H_2)^2&=\frac\hbar2(\pi^2+u^2)+\frac{\hbar^2}2,\label{eq:varH2}\\
(\Delta D)^2&=\frac\hbar2|z|^2+2\hbar^2 .\label{eq:varD}
\end{align}
Consequently, for $\Mhat=\widehat H_1^2+\widehat H_2^2+\frac12\widehat D^2$,
\begin{equation}\label{eq:sl2symbol}
m_\hbar(z)=\langle z,\hbar|\Mhat|z,\hbar\rangle=m_0(z)+\tfrac34\hbar|z|^2+2\hbar^2 .
\end{equation}
The correction on the constrained surface has a clear origin: the means of $H_1,H_2,D$ vanish there, but their fluctuations do not, and the master expectation adds them. This is the orbit-speed term once more, and at (\ref{eq:regularpoint}) its $O(\hbar)$ part is $3\hbar r^2=\frac\hbar4\Tr(KG)$, as Remark~\ref{rem:moments} predicts. The symplectic pullback is the canonical form, and (\ref{eq:sl2symbol}) converges in every $C^k$ norm on compact sets. The fixed width of the seed family is not preserved by the noncompact gauge action; the correction reflects this choice and vanishes in the limit. The quadratic master need not be gauge invariant away from $\Sigma$; it encodes the common zero set, and its characteristic image on $\Sigma$ is independent of the positive weights.

The Weyl symbol of the master operator differs from its coherent expectation:
\begin{equation}\label{eq:sl2weyl}
\Mhat=\Op\bigl(m_0+\tfrac{\hbar^2}2\bigr)\qquad\text{on }\mathcal S(\R^4).
\end{equation}
Appendix~\ref{app:symbols} verifies both symbols independently. The self-adjoint realization is fixed by the group structure.

\begin{lemma}[su(1,1) structure and self-adjointness]\label{lem:su11}
Let $\widehat G_0=\frac12(\widehat H_1-\widehat H_2)$, $\widehat G_1=\frac12(\widehat H_1+\widehat H_2)$ and $\widehat G_2=\frac12\widehat D$. Then
\begin{equation}\label{eq:su11}
[\widehat G_0,\widehat G_1]=i\hbar\widehat G_2,\qquad [\widehat G_2,\widehat G_0]=i\hbar\widehat G_1,\qquad [\widehat G_1,\widehat G_2]=-i\hbar\widehat G_0 ,
\end{equation}
and
\begin{equation}\label{eq:casimir}
\Mhat=2\bigl(\widehat G_0^2+\widehat G_1^2+\widehat G_2^2\bigr)=2\bigl(2\widehat G_0^2-\widehat Q\bigr),\qquad \widehat Q=\widehat G_0^2-\widehat G_1^2-\widehat G_2^2 .
\end{equation}
On the space of smooth vectors of the unitary representation of SL(2,$\R$) generated by the $\widehat G_\alpha$, which contains $\mathcal S(\R^4)$, $\Mhat$ is essentially self-adjoint. Its closure coincides with the operator defined by the closed form $\|\widehat H_1\psi\|^2+\|\widehat H_2\psi\|^2+\frac12\|\widehat D\psi\|^2$ and with any self-adjoint operator acting as (\ref{eq:casimir}) on smooth vectors, in particular with the spectral realization of \cite{DT3}.
\end{lemma}

\begin{proof}
The commutators follow from (\ref{eq:sl2algebra}), and (\ref{eq:casimir}) from $\widehat H_1^2+\widehat H_2^2=2(\widehat G_0^2+\widehat G_1^2)$. The quadratic constraints generate a subgroup of the metaplectic representation, whose smooth vectors include $\mathcal S(\R^4)$; since the number of copies of the defining representation is even, the representation of the subgroup is a genuine representation of SL(2,$\R$). With the anti-Hermitian generators $X_\alpha=-i\widehat G_\alpha/\hbar$, one has $\sum_\alpha X_\alpha^2=-\Mhat/(2\hbar^2)$, which is Nelson's Laplacian of the representation and is therefore essentially self-adjoint on smooth vectors \cite{Nelson}. Each of the other realizations named is a self-adjoint extension of $\Mhat$ restricted to smooth vectors, and hence equals its closure.
\end{proof}

\subsection{Actual quantum flow and the reconstructed shear}

The individual constraint flows are metaplectic: they move packet centres exactly along (\ref{eq:sl2HVF}) and transform the covariance by $V\mapsto\Lambda_gV\Lambda_g^{\rm T}$, so an invariant family must include squeezed Gaussians. For the quartic master flow, Gaussian invariance is replaced by a controlled approximation, proved in Appendix~\ref{app:propagation}.

\begin{proposition}[Master propagation]\label{prop:propagation}
Let $K\subset\R^8$ be compact and $T<\infty$. Let $z_t=\Phi^t_{m_0}(z)$ and $\Lambda_t={\rm d}\Phi^t_{m_0}|_z$. There is a normalized squeezed Gaussian $g_{z,\hbar}(t)$ centred at $z_t$, with covariance $\hbar\Lambda_t\Lambda_t^{\rm T}/2$ and an appropriate phase, such that
\begin{equation}\label{eq:propagation}
\sup_{z\in K,\ |t|\le T}\bigl\|e^{-it\Mhat/\hbar}|z,\hbar\rangle-g_{z,\hbar}(t)\bigr\|\le C_{K,T}\sqrt\hbar,\qquad 0<\hbar\le1 .
\end{equation}
At a regular $z\in\Sigma$, $z_t=z$ and $\Lambda_t=1+tL_z$.
\end{proposition}

For smooth compactly supported phase-space observables $a$, with $\psi_{z,\hbar}(t)=e^{-it\Mhat/\hbar}|z,\hbar\rangle$, boundedness of $\Op(a)$ and (\ref{eq:propagation}) imply
\begin{equation}\label{eq:egorov}
\langle\psi_{z,\hbar}(t)|\Op(a)|\psi_{z,\hbar}(t)\rangle\longrightarrow a\bigl(\Phi^t_{m_0}(z)\bigr).
\end{equation}
On $\Sigma$ the packet stays centred at a fixed classical point while its leading Gaussian shape undergoes the shear $1+tL_z$, whose image reconstructs the gauge directions. The estimate holds for fixed finite times. The Gaussian method is standard \cite{Heller,Hagedorn,Littlejohn,CombescureRobert}; the model-specific content is the identification of the covariance transport with the reconstructed gauge shear. By itself the estimate gives neither bounds on arbitrary unbounded moments of the exact state nor differentiable convergence of the exact centre map.

\subsection{A bounded-probe reconstruction of the three gauge directions}

The propagation estimate proves the fluctuation statement of Section~\ref{sec:semiclassical} for this model. Let $z$ be a regular constrained point and use the probes (\ref{eq:probes}) in eight dimensions. Since they are unitary, (\ref{eq:propagation}) gives (Appendix~\ref{app:probes})
\begin{equation}\label{eq:probebound}
\Bigl|\langle\psi_{z,\hbar}(t)|\widehat W_{z,\hbar}(\xi)|\psi_{z,\hbar}(t)\rangle-\exp\Bigl(-\frac14\,\xi^{\rm T}(1+tL_z)(1+tL_z)^{\rm T}\xi\Bigr)\Bigr|\le2C_{K,T}\sqrt\hbar ,
\end{equation}
uniformly in $\xi$ and for centres and times in the stated sets. The limiting rescaled covariance is
\begin{equation}\label{eq:sl2V}
V_z(t)=\tfrac12(1+tL_z)(1+tL_z)^{\rm T},\qquad \ddot V_z=L_zL_z^{\rm T},
\end{equation}
and $\Im\ddot V_z$ is the three-dimensional space spanned by $X_{H_1},X_{H_2},X_D$. At the point (\ref{eq:regularpoint}) the three nonzero eigenvalues of $\ddot V_z$ equal $16r^4$, so $\Tr\ddot V_z=48r^4=\Tr((KG)^2)$, in agreement with Remark~\ref{rem:moments}. The eigenvalues depend on the master normalization and the packet scale; the image is the invariant gauge subspace. No estimate on unbounded moments of the exact state is required: a packet centred at the same classical point for all $t$ encodes the unreduced gauge directions in its leading fluctuation response.

\subsection{Spectral bottom and physical-sector selection}\label{sec:bottom}

Lemma~\ref{lem:su11} makes the spectral bottom of $\Mhat$ transparent. The spectrum of $\widehat G_0=\frac14(p^2+u^2)-\frac14(\pi^2+v^2)$ is $\{\hbar k/2\}_{k\in\Z}$. In the normalization of \cite{DT3}, where the master operator is $2\Mhat$ and $\hbar=1$, the continuous part of the spectrum is $1+x^2+2k^2$, $x\ge0$, $k\in\Z$ \cite[equation~(3.34)]{DT3}. By (\ref{eq:casimir}) this is the contribution of principal-series components with $\widehat G_0=\hbar k/2$ and Casimir $\widehat Q=-\hbar^2(1+x^2)/4$, which give the generalized eigenvalues $\frac12\hbar^2(1+x^2+2k^2)$ of $\Mhat$. The discrete part, $4t(1-t)+2k^2$ with $1\le t\le|k|/2$ in the same normalization, is bounded below by $3$ \cite{DT3}. Hence
\begin{equation}\label{eq:bottom}
\inf\spec\Mhat=\tfrac12\hbar^2 ,
\end{equation}
attained at the edge of the continuous spectrum, not as an eigenvalue. The offset is of order $\hbar^2$, as Proposition~\ref{prop:general}(b) requires for the first-class zero set of this model. Subtracting it gives $\Mhat_0=\Mhat-\hbar^2/2$, whose Weyl symbol is exactly $m_0$. The scalar shift changes only an overall phase of the kinematical evolution and leaves the projective HVF, the Gaussian shear and the classical limit unchanged.

The spectral sector selected after the shift need not satisfy all intended observable relations. In \cite[equation~(3.35)]{DT3} the shifted solution space violates the classical condition $Q_3P_3=0$, with $Q_3=(O_{12}-O_{34})/2$ and $P_3=(O_{12}+O_{34})/2$, and the master operator is modified to $\Mhat-1+(Q_3P_3)^2$ in their normalization. Correct kinematical symbols and propagation are therefore insufficient to identify a physical quantum theory.

The reconstruction explains why it cannot settle this choice. Let $Y=Q_3P_3=(O_{12}^2-O_{34}^2)/4$. The two-dimensional angular-momentum identity gives the redundant relation used in \cite{DT3},
\begin{equation}\label{eq:redundant}
Y=\tfrac14\bigl(2u^2H_1-2v^2H_2-(u\cdot p+v\cdot\pi)D\bigr).
\end{equation}
For $m_\lambda=m_0+\lambda Y^2$, with $\lambda>0$ carrying the appropriate units, the zero set is unchanged and
\begin{equation}\label{eq:mlambda}
\Hess m_\lambda|_\Sigma=\Hess m_0|_\Sigma+2\lambda\,{\rm d}Y\otimes{\rm d}Y .
\end{equation}
By (\ref{eq:redundant}), ${\rm d}Y|_\Sigma$ is a combination of the three constraint gradients, so the kernel, the image and the counting of Theorem~\ref{thm:reconstruction} are unchanged on every regular stratum, while the transverse response coefficients change. Equality of the characteristic distribution does not establish equivalence of the quantum prescriptions; it is a necessary geometric comparison, to be supplemented by the physical observable representation.

Group averaging in this model is possible on a suitably chosen test space \cite{LoukoRovelli}. Noncompactness prevents a normalized Haar projection, but neither noncompactness nor nonamenability by itself rules out rigging-map averaging. The regular physical sectors and their observable representations must be compared directly with the master prescription; the different quantizations and their semiclassical requirements are compared in \cite{GambiniOlmedo}. We do not establish an equivalence of these physical constructions.

\section{Discussion}\label{sec:discussion}

\subsection{Where the classical gauge directions reside}

Squaring the constraints removes the first-order correspondence on the zero set: the master HVF vanishes there, and the master orbit through a constrained point is not the classical gauge orbit. The second-order information survives. The Hessian identifies the conormal directions, and the symplectic form maps them to the symplectic orthogonal of the zero set. Three pieces of information are thereby separated: the zero set selects the admissible configurations, the transverse Hessian specifies how departures from it are penalized, and the ambient symplectic form converts that response into the characteristic distribution and into Dirac's count of first- and second-class constraints. Positive weights change the response but not its image or the count.

This is why the trivial self-commutator of the master operator is not the relevant test of the classical constraint algebra. The relevant test is the Jordan type of the linearized master HVF, and it has two quantum counterparts. Spectrally, a first-class zero set gives an offset of order $\hbar^2$ at most (Proposition~\ref{prop:general}), while uniformly second-class directions give an offset of order $\hbar$. Dynamically, the leading fluctuations of a packet at a fixed constrained point are sheared into the gauge directions, as shown exactly in the oscillator and with bounded probes in the SL(2,$\R$) model.

There is no contradiction with the rotation (\ref{eq:LQ}) at exact physical rays. The two linearizations are taken on different spaces and at different stages of reduction: one at a physical ray at fixed $\hbar$, the other at a classical point reached through kinematical packets as $\hbar\to0$. These packets have positive master expectation at every nonzero $\hbar$. The oscillator makes this explicit: the kinematical family gives a presymplectic sphere with Hopf null circles in the limit, while projection gives a separate physical family whose symplectic form tends to the reduced form without acquiring a kernel. The quotient has already been implemented by the quantum constraint, and it is the kinematical comparison that recovers the unreduced geometry.

\subsection{Real and imaginary gauge directions}\label{sec:realimag}

Proposition~\ref{prop:kahler} gives a transparent picture of what the master flow does in the presence of a complex structure, which a family of coherent states always provides. The gauge fields $X_{C_i}$ span the characteristic distribution, the real gauge directions. Their images $JX_{C_i}=\nabla C_i$ span the normal space of $\Sigma$, the imaginary gauge directions, generated by the imaginary part of the complexified gauge group. The linearized master flow maps the second onto the first with the positive matrix $KG$, and Remark~\ref{rem:moments} shows that the first two moments of this matrix are the two semiclassical signatures of the gauge structure: the static $O(\hbar)$ offset of the master expectation, which is the orbit-speed term of \cite{Corichi2008}, and the rate at which kinematical packets are sheared.

The same pair of directions appears in group averaging, seen from the opposite side. The fibres of the averaging map, the sets of seeds that produce the same physical state, are invariant under the complexified gauge group \cite{Corichi2008}, and they are in general much larger than gauge orbits \cite{CorichiGA}. Within the coherent family of the oscillator, each averaging fibre is exactly one complexified gauge orbit: its real part is the common-phase circle of the coherent label, which on the constraint surface is the Hopf circle of Section~\ref{sec:oscillator}, and its imaginary part is the radial direction, transverse to the surface. For the coherent states of any linear action of a compact group the same holds on the whole polystable set, and within each fibre the probability that a seed is physical is maximal exactly on the gauge orbit that lies on the constraint surface \cite{CorichiDirac}. For linear first-class constraints and Gaussian seeds centred on the constraint surface, the kernel of the averaging map on displacements is exactly the sum of the real and imaginary gauge directions \cite{CorichiLinear}. Group averaging therefore \emph{identifies} the real and imaginary gauge directions, mapping the whole complexified orbit to one physical ray, while the master flow \emph{relates} them, converting the one into the other. The probability that a seed survives the projection, finally, \emph{selects} the constraint surface within each complexified orbit. In all three constructions the relevant states are kinematical. The physical ray space itself has no degenerate directions, and only before the quotient is taken can the gauge geometry be seen.

\subsection{Which constraint algebra is recovered?}\label{sec:algebra}

The reconstruction determines the characteristic distribution, the class of the constraints and the local algebra of functions vanishing on the surface, hence a geometric first-class structure, the conormal Lie algebroid of Section~\ref{sec:reconstruction}. It does not determine a preferred set of generators. Even classically, different regular defining functions generate the same foliation with different structure functions: adapted coordinates can make the constraints commute, while a physically useful non-Abelian frame does not. What is reconstructed is the equivalence class of local first-class descriptions of the same coisotropic surface. Particular generators require additional observables or geometric input; in the SL(2,$\R$) model they are supplied by the given quadratic constraints, not inferred from positivity.

Global gauge structure likewise requires further data. An integrable distribution determines connected local leaves, while a global group action, disconnected identifications, stabilizers and the topology of the leaf space are not determined locally. The theorem cannot resolve the exceptional non-Hausdorff set of the noncompact model, and the rank drop on its singular strata marks the boundary of its applicability.

\subsection{Spectral offsets, physical sectors and the variational principle}

When the master spectrum does not contain zero, the standard procedure is to subtract its minimum \cite{Thiemann,DT2}. Section~\ref{sec:offset} gives this procedure a precise diagnostic content, with two qualifications. First, the order of the offset is a global quantity: it is set by the least second-class point of the zero set, and an offset that is $o(\hbar)$ does not certify that the constraints are first class (Example~\ref{ex:magnetic}(iii)). Second, the diagnostic concerns the master function only through its principal symbol. An ordering term of order $\hbar$ that does not vanish on $\Sigma$ would shift the leading offset by $\hbar$ times its infimum over $\Sigma$, as the subprincipal term does in Melin's inequality, and must be excluded by the construction of $\Mhat$, as it is for sums of squares of self-adjoint constraint operators.

The SL(2,$\R$) model also shows the limits of any purely variational characterization of physical states. In \cite{Corichi2008} it is suggested that physical states could be defined as the minimum set of the quantum constraint function, a proposal that covers first-class constraints, the shifted master constraint and the Gupta--Bleuler treatment of second-class pairs. In SL(2,$\R$) the bottom of the spectrum is attained only at the edge of the continuous spectrum, and the sector it selects violates the classical relation $Q_3P_3=0$ \cite{DT3}. The principle therefore needs to be supplemented by the physical observable representation, as the reconstruction of the gauge distribution does. Equality of zero sets, of characteristic distributions and of spectral bottoms are all necessary checks, but none of them is sufficient.

\subsection{What is needed for general relativity?}

Canonical gravity is a demanding application, because its constraints generate deformations of a spatial hypersurface. After internal gauge reduction, the Hamiltonian and spatial diffeomorphism constraints satisfy, schematically and up to a convention-dependent sign,
\begin{equation}\label{eq:hypersurface}
\{H[N],H[N']\}=V\bigl[q^{ab}(N\partial_bN'-N'\partial_bN)\bigr],
\end{equation}
together with the brackets involving the diffeomorphism constraint $V[\vec N]$. The inverse spatial metric is essential to the interpretation as hypersurface deformations \cite{HKT}. A characteristic distribution alone does not specify which generator is normal, which is tangential, or which field is the spatial metric, so recovering (\ref{eq:hypersurface}) needs more than a first-class test.

A positive master functional built from smeared constraints would have a Hessian of the factorized form (\ref{eq:quadL}), and with controlled functional analysis its image could identify the infinitesimal gauge distribution. A result for gravity would require function spaces, domains, regularity of the constraint map and control of the range of the Hessian; the finite-dimensional identity between the range of a symmetric map and the annihilator of its kernel does not transfer directly to weakly symplectic infinite-dimensional phase spaces. The spectral test would also need an infinite-dimensional version, in which $\Tr_+L$ is replaced by a renormalized trace. The state family must encode spatial locality and a candidate metric, and the regulator or continuum limit must preserve the relevant ranks and brackets. Master-constraint analyses of Maxwell theory and linearized gravity already show that squares of operator-valued distributions require care \cite{DT4}. The present results supply local diagnostics for controlled finite-dimensional approximations, such as regularized or truncated models; they do not establish a continuum constraint algebra for loop quantum gravity.

These considerations refine the role of anomaly freedom. The identity $[\Mhat,\Mhat]=0$ is automatic. A satisfactory classical correspondence requires instead a coisotropic limiting surface, the correct characteristic distribution, an appropriate physical observable representation and, through (\ref{eq:hypersurface}), a physically identified frame with the intended metric-dependent structure functions. These are distinct checks, and the first two are precisely what the results of Sections~\ref{sec:reconstruction} and \ref{sec:semiclassical} can test. In a finite-dimensional approximation, constraints that fail to close on the limiting zero set are second class there, and if this happens at every point the master offset is of order $\hbar$. A failure of closure that vanishes in the classical limit, a genuine quantum anomaly of relative order $\hbar$, behaves differently: for a central anomaly the offset is only of order $\hbar^2$, the order of ordering effects, so the order of the offset does not detect it, while the master dynamics does. This case is analysed in \cite{CorichiAnomalies}.

\subsection{Physical information beyond the gauge distribution}

The redundant-constraint modification of Section~\ref{sec:bottom} preserves the zero set and the gauge distribution but changes the spectral selection. The reconstructed gauge geometry therefore cannot decide which quantum prescription carries the intended physical content; physical inner products, sectors and observables must also be compared. Conversely, equal reduced theories need not select a unique unreduced gauge presentation, and different kinematical extensions can describe the same reduced observables. A quantum-fundamental viewpoint should not demand uniqueness where the classical description itself allows freedom. The question is whether a proposed semiclassical interpretation has a consistent first-class geometry and reproduces the relevant physical observables, with any spacetime interpretation stated explicitly.

\section{Conclusions}\label{sec:conclusions}

We have studied how an unreduced classical constrained geometry, and the class of its constraints, emerge from quantum master data. For a Morse--Bott master function, the linearized HVF is the Hamilton map of the master function: its image is the symplectic orthogonal of the zero set, and its Jordan structure counts the first- and second-class constraints. This information is accessible to quantum theory in three ways, which require increasing input. The bottom of the master spectrum needs only the family of master operators: it is $(\hbar/2)\Tr_+L$ for quadratic masters, at most of order $\hbar^2$ for every first-class zero set, and of order $\hbar$ when second-class directions are present everywhere, although it detects only the least second-class point of the zero set. The static offset of the master expectation needs a coherent family, and the covariance shear needs in addition the master unitary; the two are the first and second moments of a single matrix.

The oscillator gives exact master evolution, an exact realization of the covariance shear, and a physical coherent family with the correct reduced symplectic form. The SL(2,$\R$) model gives a non-Abelian reconstruction, exact coherent and Weyl master symbols, essential self-adjointness of the master operator, its spectral bottom $\hbar^2/2$, and a finite-time propagation estimate realizing the fluctuation diagnostic.

The geometric difference between classical and quantum constraints is thereby clarified rather than removed. Exact physical quantum states describe an already reduced theory, and classical gauge directions arise in the response of kinematical semiclassical states near a limiting constraint surface. Reconstructing those directions, and the class of the constraints, is a controlled local result, while selecting physical sectors and identifying the spacetime constraint frame require additional input. Much remains to be done before these ideas can be applied to gravity itself. We hope that they provide a precise starting point for asking how a fundamental quantum theory recovers the constrained geometry of gravity.

\section*{Acknowledgments}
I thank Steph and Fide for discussions and support. This work received no external funding.

\section*{Data availability statement}
All data that support the findings of this study are included within the article. A script that verifies the closed forms and the numbers quoted in the text, and a script that generates Figures~\ref{fig:shear} and \ref{fig:osc}, are provided as supplementary material.

\appendix

\section{Proof of Theorem~\ref{thm:reconstruction}: complements}\label{app:reconstruction}

Let $H:T_z\Gamma\to T_z^*\Gamma$ be the symmetric Hessian map with $\ker H=T_z\Sigma$. Then $H(T_z\Gamma)\subset(T_z\Sigma)^\circ$, since $H(\delta z,t)=H(t,\delta z)=0$ for $t\in T_z\Sigma$. Both spaces have dimension $\dim\Gamma-\dim\Sigma$, so they coincide. The map $\omega^\flat$ carries $(T_z\Sigma)^\omega$ onto $(T_z\Sigma)^\circ$, hence $\Pi^\sharp H(T_z\Gamma)=(T_z\Sigma)^\omega$, without choosing constraints or coordinates.

For part (c), note that $H\ge0$ because $z$ is a minimum of $m$. In the normal form (\ref{eq:williamson}) of $H$, the coordinates $x_k$ of the nilpotent blocks span $\ker H\cap\Im\Pi H=T_z\Sigma\cap(T_z\Sigma)^\omega$, the rotation planes span a symplectic complement of $T_z\Sigma$ inside $(T_z\Sigma)^\omega+T_z\Sigma$, and the remaining coordinates span a symplectic subspace of $T_z\Sigma$ on which $H$ vanishes. In Dirac's terms these are, respectively, the first-class directions, the second-class pairs and the physical directions at $z$.

When $D=\ker\omega_\Sigma$ has constant rank, it is a smooth distribution. For $Y,Z$ in it, ${\rm d}\omega_\Sigma=0$ and Cartan's formula give $\mathcal L_Y\omega_\Sigma=0$ and $\iota_{[Y,Z]}\omega_\Sigma=\mathcal L_Y\iota_Z\omega_\Sigma-\iota_Z\mathcal L_Y\omega_\Sigma=0$. Frobenius' theorem gives local leaves, along which $\omega_\Sigma$ is invariant; this is the condition for the form to descend to a smooth quotient. In the coisotropic case $D=\Im L$ has constant rank $\codim\Sigma$.

For structure functions, use local coordinates $(x,C)$ adapted to $\Sigma$ and let $b_{ij}(x,C)=\{C_i,C_j\}$. Coisotropy gives $b_{ij}(x,0)=0$, and Hadamard's lemma,
\begin{equation}\label{eq:hadamard}
b_{ij}(x,C)=C_k\int_0^1\frac{\partial b_{ij}}{\partial C_k}(x,sC)\,{\rm d}s ,
\end{equation}
gives smooth local $F_{ij}{}^k$, which are not unique. The quotient $I_\Sigma/I_\Sigma^2$ identifies functions with the same conormal first derivative, and its bracket is well defined because $\{I_\Sigma^2,I_\Sigma\}\subset I_\Sigma^2$. With our convention, the anchor $[f]\mapsto-X_f|_\Sigma$ is a Lie-algebra homomorphism, while $[f]\mapsto X_f|_\Sigma$ is an antihomomorphism.

For independent constraints at $C_i=0$, differentiating $m=K^{ij}C_iC_j/2$ twice leaves only $K^{ij}{\rm d}C_i\otimes{\rm d}C_j$. Its kernel is $\bigcap_i\ker{\rm d}C_i=T\Sigma$ because $K$ is positive definite, and its rank is the number of constraints. Applying $\Pi^\sharp$ gives (\ref{eq:quadL}), and ${\rm d}C_j(X_{C_k})=\{C_j,C_k\}$ gives (\ref{eq:LDirac}). Along the flow of an observable $O$, the second derivative of $m$ at a point of $\Sigma$ is $\Hess m(X_O,X_O)$, since the acceleration term multiplies ${\rm d}m=0$. With our convention this equals $\{O,\{O,m\}\}$, which proves (\ref{eq:doublebracket}) and, by positivity, its equivalence to the tangency of $X_O$.

\section{Proof of Proposition~\ref{prop:general}}\label{app:general}

For $\psi\in\mathcal S(\R^n)$ and a symbol $f$ of polynomial growth, $\ip{\psi}{\Op(f)\psi}=\int f\,W_\psi$, where $W_\psi$ is the Wigner function of $\psi$. Since $\mathcal S(\R^n)$ lies in the form domain of $\Mhat_\hbar$, $\inf\spec\Mhat_\hbar\le\ip{\psi}{\Op(m)\psi}+\hbar^2\ip{\psi}{\Op(m_2)\psi}$ for every normalized Schwartz $\psi$. For both families of trial states below, $\ip{\psi}{\Op(m_2)\psi}$ is bounded uniformly in $0<\hbar\le1$, so it suffices to bound the first term.

\emph{(a) Gaussian trial states.} Let $z\in\Sigma$ and $H=\Hess_zm\ge0$. Choose a linear symplectic map $S$ that brings $\frac12\zeta^{\rm T}H\zeta$ to the normal form (\ref{eq:williamson}), with $r$ rotation blocks, $s$ nilpotent blocks and $n-r-s$ null pairs. Let $\varphi$ be the product Gaussian whose Wigner function is a centred normal density with covariance $\frac\hbar2$ in each coordinate of the rotation blocks and of the null pairs, and with variances $\frac12\hbar^{1+2\delta}$ in $\xi_k$ and $\frac12\hbar^{1-2\delta}$ in $x_k$ for the nilpotent blocks, $0<\delta<\frac14$. Let $\psi$ be $\varphi$ transformed by the metaplectic operator of $S$ and displaced to $z$. Its Wigner function is $W_\varphi(S^{-1}(\cdot-z))$, a positive Gaussian density, by the metaplectic covariance of the Weyl calculus \cite{Folland}. Taylor's theorem with polynomial growth gives
\[
m(z+S\zeta)=\tfrac12\zeta^{\rm T}S^{\rm T}HS\,\zeta+c_3(\zeta)+R_4(\zeta),\qquad|R_4(\zeta)|\le C|\zeta|^4(1+|\zeta|^N),
\]
with $c_3$ a cubic form. Under the centred Gaussian, $c_3$ averages to zero, the quadratic part averages to $\frac\hbar2\sum_j\mu_j+\frac s4\hbar^{1+2\delta}$, and $R_4$ averages to at most $C'\hbar^{2-4\delta}$, since the largest variance is $\frac12\hbar^{1-2\delta}$. Choosing $\delta=\frac16$ gives (\ref{eq:generalA}).

\emph{(b) Lagrangian trial states.} Near a point $z_0$ where $\Sigma$ is coisotropic of codimension $k$, the coisotropic normal form gives Darboux coordinates $(Q,P,x,y)$, $Q,P\in\R^k$, in which $\Sigma=\{P=0\}$ \cite{Gotay}. The submanifold $\Lambda=\{P=0,\ y=0\}$ is Lagrangian and contained in $\Sigma$. After a linear symplectic change of the original coordinates, implemented exactly by a metaplectic operator, $T_{z_0}\Lambda$ is transverse to the fibres $\{q={\rm const}\}$, so near $z_0$ the submanifold $\Lambda$ is the graph $\{(q,\nabla\phi(q)):q\in U\}$ of an exact form, with $\phi$ smooth on an open set $U$, which we extend to a smooth function with bounded derivatives on $\R^n$. Let $a\in C_c^\infty(U)$ be \emph{real}, with $\|a\|=1$, and $\psi=a\,e^{i\phi/\hbar}$. With $X=\frac12(x+y)$, $w=x-y$ and $\Phi(x,y)=\int_0^1\nabla\phi(y+s(x-y))\,{\rm d}s=\nabla\phi(X)+O(|w|^2)$,
\[
\ip{\psi}{\Op(m)\psi}=\frac1{(2\pi\hbar)^n}\int a(X+\tfrac w2)\,a(X-\tfrac w2)\,e^{iw\cdot\xi/\hbar}\,m\bigl(X,\xi+\Phi(X+\tfrac w2,X-\tfrac w2)\bigr)\,{\rm d}w\,{\rm d}\xi\,{\rm d}X .
\]
The linear term of $\Phi$ in $w$ vanishes because the midpoint expansion is even in $w$. Each power of $w$ is traded for a factor $\hbar\,\partial_\xi$ by integration by parts, so replacing $\Phi$ by $\nabla\phi(X)$ changes the integral by $O(\hbar^2)$. In the remaining integral we set $\xi=\hbar\eta$ and write it as $\int m(X,\nabla\phi(X)+\hbar\eta)\,F(X,\eta)\,{\rm d}\eta\,{\rm d}X$, where $F(X,\eta)=(2\pi)^{-n}\int a(X+\frac w2)a(X-\frac w2)e^{iw\cdot\eta}\,{\rm d}w$ is rapidly decreasing in $\eta$, has $\int F\,{\rm d}\eta=a(X)^2$, and is even in $\eta$ because $a$ is real. Expanding in $\hbar\eta$, the zeroth-order term $\int m(X,\nabla\phi(X))\,a(X)^2\,{\rm d}X$ vanishes because $\Lambda\subset\Sigma$; the first-order term vanishes because $F$ is even in $\eta$; and the remainder is $O(\hbar^2)$ because $F$ has finite moments of all orders. This proves (\ref{eq:generalB}).

The proof of (b) is where the class of the constraints enters: a Lagrangian submanifold of $\Gamma$ contained in $\Sigma$ exists near $z_0$ if and only if $\Sigma$ is coisotropic near $z_0$, since a maximal isotropic subspace of $T_z\Sigma$ has dimension $n-\frac12\rank L_z^2$. If the amplitude $a$ is allowed to be complex, the first-order term becomes $\hbar\int{\rm Im}(\bar a\nabla a)\cdot\partial_\xi m\,{\rm d}X$, which vanishes as well, since $\partial_\xi m=0$ on $\Sigma$.

\section{Coherent and Weyl symbols in the non-Abelian model}\label{app:symbols}

For one Gaussian coordinate with mean $x$ and variance $s^2=\hbar/2$, ${\rm Var}(x^2)=4s^2x^2+2s^4$. The four variables entering $H_1$ commute and are independent in the product packet, and summing their variances with the factor $1/4$ gives the $H_1$ variance; the $H_2$ computation is identical.

For one mode, $B=(QP+PQ)/2$ has mean $qp$ and variance $\hbar(q^2+p^2)/2+\hbar^2/2$. The constant follows from $B=i\hbar(\hat a^{\dagger2}-\hat a^2)/2$ at zero centre, whose vacuum variance is $\hbar^2/2$. Four independent modes, with the signs in $D$, give ${\rm Var}\,D=\hbar|z|^2/2+2\hbar^2$, and combining the variances proves (\ref{eq:sl2symbol}).

For the Weyl symbol, $H_1\star H_1=H_1^2$ and $H_2\star H_2=H_2^2$, since each depends only on mutually commuting canonical variables. In one mode $(qp)\star(qp)=(qp)^2+\hbar^2/4$, so $D\star D=D^2+\hbar^2$ in four modes, which proves (\ref{eq:sl2weyl}). Independently, Gaussian averaging of a Weyl symbol is $e^{(\hbar/4)\Delta}$. With $\Delta m_0=3|z|^2$ and $\Delta^2m_0=48$, $e^{(\hbar/4)\Delta}(m_0+\hbar^2/2)=m_0+\frac34\hbar|z|^2+2\hbar^2$, in agreement with (\ref{eq:sl2symbol}).

\section{Finite-time Gaussian propagation}\label{app:propagation}

We give the elementary polynomial-residual proof of Proposition~\ref{prop:propagation}, an instance of the standard coherent-packet construction \cite{Heller,Hagedorn,Littlejohn,CombescureRobert}.

The classical flow is complete. Conservation of $m_0$ bounds $H_1,H_2,D$ along a trajectory, and the equation
\begin{equation}\label{eq:classflow}
\dot z=2H_1X_{H_1}+2H_2X_{H_2}+DX_D
\end{equation}
is linear in $z$ with bounded time-dependent coefficients, so Gronwall's inequality excludes finite-time escape. For initial centres in $K$, trajectories with $|t|\le T$ stay in a common compact set. Their derivatives $\Lambda_t$ solve $\dot\Lambda_t=\Pi\Hess m_0(z_t)\Lambda_t$ with $\Lambda_0=1$ and are uniformly bounded.

Let $\varphi_\hbar(x)=(\pi\hbar)^{-1}\exp(-|x|^2/(2\hbar))$ on $\R^4$, let $\mathcal D(z_t)$ be the Weyl displacement operator and $\mathcal U(\Lambda_t)$ the continuous metaplectic lift starting at the identity, and define
\begin{equation}\label{eq:ansatz}
g_{z,\hbar}(t)=e^{i\gamma(t)/\hbar}\,\mathcal D(z_t)\,\mathcal U(\Lambda_t)\,\varphi_\hbar ,
\end{equation}
where, with $z_t=(q_t,p_t)$,
\begin{equation}\label{eq:phase}
\dot\gamma=\tfrac12\bigl(p_t\cdot\dot q_t-q_t\cdot\dot p_t\bigr)-m_0(z_t)-\hbar^2/2,\qquad \gamma(0)=0 .
\end{equation}
Differentiating this ansatz reproduces exactly the constant, linear and quadratic terms of the Weyl-ordered Taylor expansion of $m_0+\hbar^2/2$ at $z_t$, and its covariance is $\hbar\Lambda_t\Lambda_t^{\rm T}/2$.

Since $m_0$ is a quartic polynomial, the remainder consists of a cubic and a quartic Weyl polynomial in centred phase-space variables. On a Gaussian with bounded squeezing, a centred Weyl polynomial of degree $k$ has norm bounded by a constant times $\hbar^{k/2}$, as follows from metaplectic covariance and the scaling $x=\sqrt\hbar\,y$. The coefficients are uniformly bounded on the compact trajectory set, so the residual $r(t)=(i\hbar\partial_t-\Mhat)g_{z,\hbar}(t)$ obeys
\begin{equation}\label{eq:residual}
\|r(t)\|\le C_{K,T}\bigl(\hbar^{3/2}+\hbar^2\bigr).
\end{equation}
The Gaussians are Schwartz vectors, hence in the domain of $\Mhat$ (Lemma~\ref{lem:su11}). Duhamel's formula and unitarity give
\begin{equation}\label{eq:duhamel}
\bigl\|e^{-it\Mhat/\hbar}|z,\hbar\rangle-g_{z,\hbar}(t)\bigr\|\le\frac1\hbar\int_0^{|t|}\|r(s)\|\,{\rm d}s\le C'_{K,T}\sqrt\hbar .
\end{equation}
At a regular constrained centre $z_t=z$, so the Hessian is constant and $\Lambda_t=e^{tL_z}=1+tL_z$. For bounded smooth Weyl observables the norm error controls expectation errors, and Gaussian localization gives the classical limit. No estimate on unbounded moments is inferred from the norm bound.

\section{Exact oscillator formulas}\label{app:oscillator}

In a two-mode coherent state the total number is Poisson distributed with mean $R=S/\hbar$. Hence $\ev{\Chat_\hbar}=\hbar\nu(R+1)-E$ and ${\rm Var}\,\Chat_\hbar=\hbar^2\nu^2R$, and $\ev{\Chat_\hbar^2}=\ev{\Chat_\hbar}^2+{\rm Var}\,\Chat_\hbar$ proves (\ref{eq:oscsymbol}).

For the centre, $e^{it\Mhat_\hbar/\hbar}\hat a_je^{-it\Mhat_\hbar/\hbar}=e^{-it[\mathcal E(\hat n+1)-\mathcal E(\hat n)]/\hbar}\hat a_j$. Acting with $\hat a_j$ on the coherent state leaves the Poisson weights of the remaining total number $k$ unchanged, so
\begin{equation}\label{eq:centresum}
\ev{\hat a_j}_t=\alpha_j\,e^{-R}\sum_{k=0}^\infty\frac{R^k}{k!}\,e^{-it[\mathcal E(k+1)-\mathcal E(k)]/\hbar}.
\end{equation}
With $[\mathcal E(k+1)-\mathcal E(k)]/\hbar=\hbar\nu^2(k+\frac32)-\nu E$, summing the exponential series gives (\ref{eq:osccentre}). Its squared modulus is $|z_j|^2\exp[2S(\cos(t\hbar\nu^2)-1)/\hbar]$. The sum of uncentred quadrature second moments is $2S+2\hbar$, and subtracting the squared means gives (\ref{eq:oscspread}); Taylor expansion for bounded $t$ gives the uniform $O(\hbar)$ spread.

In the same way, $e^{it\Mhat_\hbar/\hbar}\hat a_j\hat a_ke^{-it\Mhat_\hbar/\hbar}=e^{-it[\mathcal E(\hat n+2)-\mathcal E(\hat n)]/\hbar}\hat a_j\hat a_k$, with $[\mathcal E(k+2)-\mathcal E(k)]/\hbar=2\hbar\nu^2(k+2)-2\nu E$, which gives (\ref{eq:pairmoment}). For the covariance, write $\varphi=t\hbar\nu^2$, so that $R\varphi=t\nu^2S$ and $R\varphi^2=t^2\hbar\nu^4S$. Then
\begin{equation}\label{eq:covexpansion}
\begin{aligned}
{\rm Cov}(\hat a_j,\hat a_k)&=\alpha_j\alpha_k\,e^{2it\nu E}\Bigl[e^{-4it\hbar\nu^2+R(e^{-2i\varphi}-1)}-e^{-3it\hbar\nu^2+2R(e^{-i\varphi}-1)}\Bigr]\\
&=-\frac{z_jz_k}\hbar\,e^{2it\nu(E-\nu S)}\bigl(it\hbar\nu^2+t^2\hbar\nu^4S\bigr)+O(\hbar),\\
{\rm Cov}(\hat a_j^\dagger,\hat a_k)&=\bar\alpha_j\alpha_k\bigl(1-e^{2R(\cos\varphi-1)}\bigr)=\frac{\bar z_jz_k}\hbar\,t^2\hbar\nu^4S+O(\hbar),
\end{aligned}
\end{equation}
which is the input to Proposition~\ref{prop:osc-cov}. The symmetrized quadrature covariance follows from $\hat a_j=(\widehat Q_j+i\widehat P_j)/\sqrt{2\hbar}$; for instance $\hbar^{-1}{\rm Var}\,\widehat Q_j=\frac12+{\rm Re}\,{\rm Cov}(\hat a_j,\hat a_j)+{\rm Cov}(\hat a_j^\dagger,\hat a_j)$.

Projection onto total number $N$ gives a vector proportional to $(\alpha_1\hat a_1^\dagger+\alpha_2\hat a_2^\dagger)^N|0\rangle$. In the chart $\alpha_1\neq0$ its ray depends only on $w=\alpha_2/\alpha_1$, and the holomorphic representative $(\hat a_1^\dagger+w\hat a_2^\dagger)^N|0\rangle/\sqrt{N!}$ has norm $(1+|w|^2)^{N/2}$. Pulling back (\ref{eq:Omega}) gives $\hbar_NN\,\omega_{\rm FS}$.

\section{Covariance reconstruction and bounded probes}\label{app:probes}

Let $V_0$ be positive definite and $B=LV_0^{1/2}$. Then $LV_0L^{\rm T}=BB^{\rm T}$ has kernel $\ker B^{\rm T}$ and image $\Im B=\Im L$. If $L^2=0$, expanding $(1+tL)V_0(1+tL)^{\rm T}$ gives (\ref{eq:Vddot}). This holds in any fixed tangent frame, with transposes read as dual maps.

For normalized $\psi$, $g$ and a bounded operator $W$ of norm one,
\begin{equation}\label{eq:probediff}
|\ip{\psi}{W\psi}-\ip{g}{Wg}|\le2\|\psi-g\| ,
\end{equation}
uniformly for any family of such operators, including the $\hbar$-dependent probes (\ref{eq:probes}). A Gaussian centred at $z$ with covariance $\hbar V(t)$ has probe expectation $\exp[-\xi^{\rm T}V(t)\xi/2]$. Applying (\ref{eq:propagation}) proves (\ref{eq:probebound}). One takes the limit of the bounded expectations first and then defines the covariance as the Hessian of $-\log\chi_z$ at $\xi=0$, so no interchange of limits is needed.

At the point (\ref{eq:regularpoint}), $AA^{\rm T}={\rm diag}(2r^2,2r^2,4r^2)$, so the nonzero eigenvalues of the Hessian $A^{\rm T}KA$ all equal $4r^2$. Since $\Pi$ is orthogonal in the canonical frame, $L=\Pi A^{\rm T}KA$ has the same singular values, and $LL^{\rm T}$ has eigenvalue $16r^4$ with multiplicity three and $0$ with multiplicity five. Its image is the reconstructed gauge space. In the notation of Remark~\ref{rem:moments}, $G=AA^{\rm T}$ in this frame and $KG=4r^2\,\mathbf 1_3$, so $\frac\hbar4\Tr(KG)=3\hbar r^2$ and $\Tr((KG)^2)=48r^4$.

\end{document}